\documentclass[twocolumn, amssymb,nobibnotes,aps,prb,nopacs]{revtex4}

\usepackage{amsmath}
\usepackage{amssymb}
\usepackage{graphicx}
\usepackage{subfig}
\newcommand \dd[1]  { \,\textrm d{#1} }   
\usepackage[justification=RaggedRight,font=footnotesize]{caption}
\usepackage{multirow}

\begin{document}
\title{A study of the stability properties of Sagdeev solutions 
\\ in the ion-acoustic regime using kinetic simulations}
\author{S. M. Hosseini Jenab\footnote{Email: Mehdi.Jenab@umu.se}}
\author{F. Spanier \footnote{Email: Felix@fspanier.de}}
\author{G. Brodin  \footnote{Email: Gert.Brodin@umu.se}}
\affiliation{Centre for Space Research, North-West University,
Private Bag X6001, Potchefstroom 2520, South Africa}
\affiliation{Department of Physics, 
Ume\aa University, 
SE-90187 Ume\aa, 
Sweden}
\date{\today}

\begin{abstract}
The Sagdeev pseudo-potential approach has been employed
extensively in theoretical studies to determine
large-amplitude (fully) nonlinear solutions in a variety of multi-species plasmas. 
Although these solutions are repeatedly considered as solitary waves (and even solitons), 
their temporal stability has never been proven.
In this paper,
a numerical study of the Vlasov-Poisson system 
is made to follow their temporal evolution
in the presence of numerical noise
and thereby test their long-time propagation stability.
Considering the ion-acoustic regime, both constituents of the plasma, i.e. electrons and ions
are treated following their distribution functions
in these set of fully-kinetic simulations.
The findings reveal that 
the stability of Sagdeev solution depends on a combination of two parameters, 
i.e. velocity and trapping parameter.
It is shown that there exists a critical value of trapping parameter for both fast and slow solutions
which separates the stable from unstable solutions.
In case of stable solutions, 
it is shown that these nonlinear structures
can propagate for long periods, which
confirms their status as solitary waves.
Stable solutions are reported for both Maxwellian and Kappa distribution functions.
For unstable solutions, it is demonstrated that 
the instability causes the Sagdeev solution 
to decay by emitting ion-acoustic wave-packets
on its propagation trail.
The instability is shown to take place in a large range of velocity
and even for Sagdeev solutions with velocity much higher than ion-sound speed.
Besides, in order to validate our simulation code two precautionary measures are taken.
Firstly, the well-known effect of the ion dynamics on a stationary electron hole solution
is presented as a benchmarking test of the approach. 
Secondly, In order to verify the numerical accuracy of the simulations,
the conservation of energy and entropy are presented.

\end{abstract}
\maketitle

\section{Introduction} \label{Sec_Introduction}

\subsection{Theoretical study of ion-acoustic solitary waves}
The study of nonlinear waves plays a crucial role in science  as 
we know it today\cite{Scott2007nonlinear}, and the field of nonlinear physics
has opened numerous research frontiers and 
technological applications. 
In the heart of this research field stands the crucial question of finding fundamental 
nonlinear modes, in particular \emph{solitary waves}\cite{Scott2006encyclopedia}. 
By definition, a solitary wave can propagate
without any temporal evolution in shape or
size \cite{scott2003nonlinear}.
Discovered accidentally in 1834 by John Scott Russell \cite{Russel1845}
in ``the happiest day of his life''\cite{Russel1865},
these waves have defied the narrative of 
physics since their discovery and pushed physicists to a new realm, i.e.
nonlinear physics.

The study of solitary waves in plasmas started 
with the work of Washimi and Taniuti \cite{Washimi1966996}, 
who successfully derived the \emph{Korteweg–de Vries (KdV) equation} \cite{Jager2006}
for ion-acoustic waves in the nonlinear regime utilizing the well-known 
``reductive perturbation method''
\cite{taniuti1968reductive,kakutani1968reductive}. 
Although effective in deriving evolution equations for weakly nonlinear systems, 
this method cannot be applied beyond the small amplitude limit.

The Sagdeev pseudo-potential method~\cite{sagdeev1966cooperative}
(also known as mechanical-motion analog)
offers an alternative approach to the problem
of finding solitary waves, 
and it overcomes the limitation of
small amplitude approximations. 
However, it does not provide information about the temporal evolution.
Since it does not yield to any dynamical 
equation for the solitary waves,
it cannot predict if the solutions
are able to keep their features and shapes for a long (theoretically infinite) time propagation. i.e. if they are stable. 
In other words, this method provides nonlinear solutions
for a plasma which can be considered as candidates for solitary waves.

The method basically alters the Poisson's equation 
in order to rearrange it as a general
energy equation of the form, 
\begin{equation}
    \frac{1}{2} \Big(\frac{d \phi}{d x}\Big)^2 + S(\phi) = 0,
    \label{Eq_Sagdeev}
\end{equation}
which can be compared to a general energy conservation equation. 
In Eq.~\ref{Eq_Sagdeev}, the first term plays the role of kinetic energy and 
$S(\phi)$ is the so called \emph{Sagdeev pseudo-potential}.
If we consider $\phi$ as a position and $x$ as time,
this represents a classical particle 
of unit mass  moving
in a one-dimensional potential $S(\phi)$,  as is well-known. 
This method has been adopted to numerous plasma environments,
such as dusty plasmas \cite{verheest1992nonlinear}, 
electron-positron-ion plasmas \cite{mahmood2008ion}, 
magnetized plasmas\cite{sultana2010oblique}
and magnetospheric plasmas\cite{berthomier1998solitary}.

In this context,
three concepts should be recognized 
and defined properly to avoid misunderstanding:
\begin{itemize}
    \setlength\itemsep{0.0em}
    \item Sagdeev solutions
    \item solitary waves
    \item solitons.
\end{itemize}
Sagdeev solutions are localized solutions (pulses) to Eq.~\ref{Eq_Sagdeev}
(i.e. solutions such that $\phi\rightarrow 0$ and $\dd \phi / \dd x \rightarrow 0$ when $x\rightarrow \pm \infty$),
which might or might not be solitary waves or solitons. 
To prove that they are true solitary waves,
one needs to demonstrate the stability against small perturbations, 
either by analytical means or by confirming that
the pulse shape remains fixed in a long-term simulation.
In other words, their stability should be tested 
against the inherit numerical noise of the numerical method.
On the other hand, solitons are solitary waves which can survive 
(mutual) collisions without change in shape. 
For a Sagdeev solution to be a soliton, it must satisfy 
both conditions, e.g. stability during the long-time propagation and robustness
against mutual collisions. 
For instance, it is shown that Sagdeev solutions can not survive collisions
in pair plasmas based on Vlasov simulations\cite{eliasson2005solitary}.

It should be noted that both the reductive perturbation theory
and the Sagdeev pseudo-potential approach
have mostly been developed within the fluid framework, i.e. ignoring kinetic effects.
They relay on the number density (the first moment of the distribution function)
as the starting step of the dynamical model.
The complex dependence of dynamics on the distribution function is left out of the model.
However, although the Sagdeev approach has often been used within a fluid framework, 
it does not necessarily need to be restricted in this way.
A few theoretical attempts have been made to add the 
kinetic effects such as trapping to the Sagdeev solutions \cite{sayal1993study}.

\subsection{Simulation study of electron holes}
Alongside the theoretical attempts to predict the shape solitary waves and their stability, 
there exist a line of research based on simulation focusing on the property and stability of phase space holes.
Since solitary waves are always accompanied by an  hole in the phase space (in study frame of kinetic theory) 
hence these two line of research are closely related.
In case of ion-acoustic regime this is
called ``coupled states of electron holes and ion-acoustic solitons''.
However, a fast enough electron hole can propagate without causing 
a strong reaction from ions
hence it will not excite an ion-acoustic solitary waves.
An electron hole consists of electrons resonating with 
the potential well of the propagating solitary waves (also known as trapped particles).
Over the years, some approaches have been employed as the ``generation mechanism''
to produce electron holes and hence solitary waves including:
\begin{enumerate}
\setlength\itemsep{0.0em}
    \item chain formation process (disintegration)\cite{Zabusky1965,Kakad2013,Kakad2014nonlinear,Sharma2015,Qi20153815}      
    \item electrostatic streaming instabilities \cite{califano2007electrostatic,silin2007instabilities,califano2005electron}
    \item nonlinear Landau damping \cite{manfredi1997long,brunetti2000asymptotic,abbasi2007vlasov}
    \item chirp-driving (external driving) mechanism \cite{bertsche2003direct,trivedi2016chirp,peinetti2005numerical}
    \item non-periodic plasmas\cite{fijalkow2003phase}
    \item random phase approach \cite{pezzi2014kinetic}
    \item transient approach \cite{zhou2016plasma}.
\end{enumerate}
These approaches develop the solitary waves and its corresponding electron hole
naturally and during the temporal evolution of the simulation.
They basically start from an unstable initial condition and reach a steady state which is a
subclass of Bernstein-Greene-Kruskal (BGK) modes\cite{bernstein1957exact} 
However, one should note that solitary waves and their accompanying holes coming
from these mechanisms are not necessarily equal to one another. 
In other words, they can belong to different subclasses of BGK modes.
The above list refers to just the recent papers on these mechanisms and obviously there
are many other works and research areas focusing on different aspects of 
electron holes and solitary waves\cite{hutchinson2017electron}.

The family of solutions generated through the Sagdeev method 
does not generally correspond to asymptotic solutions in the dynamic evolution. 
Hence electron holes generated by the Sagdeev method may or may not be stable,
depending on the parameters.
Furthermore, these is no guarantee that the BGK modes created in the 
different generation mechanisms can be reproduced by Sagdeev approach.

To give a few example of other related topics, 
it is reported by Saeki and Genma\cite{saeki1998electron} that 
ion dynamics cause an electron hole to split into to oppositely-propagating holes if 
its velocity is in order or slower than ion-acoustic speed.
We have adopted these well-known phenomena as the benchmark of our simulation code.

It is also can be generally assumed that any large hole
can break up into smaller holes in a process called ``chain formation''.
It is shown that the pulses produced by this process 
are truly solitary waves via long-time simulation\cite{jenab2016IASWs}
and solitons by focusing on their mutual collisions\cite{hosseini2017simulation,jenab2017fully,hosseini2017kinetic} 
in the fully kinetic regime.

There has been a strong line of research on the standing electron holes and their
dynamics and interaction with ions and each other by focusing on their merging and
break-up process\cite{eliasson2005dynamics,cheng2015energy}.
The stability of electron holes is also considered from general point of view
by approaching them as BGK modes\cite{manfredi2000stability, demeio1991numerical}.

\subsection{The scope of this investigation}
The question of our study is:
\begin{center}
Are ``Sagdeev solutions'' equal to ``solitary waves''?
\end{center}
We will provide an answer to this question in \emph{electrostatic fully-kinetic ion-acoustic} regime.
To do so, we need to study the temporal behavior of Sagdeev solutions and their corresponding electron holes
in order to establish their long-time stability.

Here, we are reporting the results in the ion-acoustic regime, 
i.e. considering electron-ion plasmas. 
Fully-kinetic ion-acoustic regime refers to the fact that
the dynamics of both electrons and ions
are followed by the Vlasov equation.
Two types of background distribution functions have been considered; the 
Maxwellian and the Kappa distribution function


We proceed as follows:
Firstly, the pulse 
profile is determined numerically based on the Sagdeev pseudo-potential approach
while accounting for trapped and reflected particles (see Sec.\ref{SubSec_Equations_Sagdeev}).   
Secondly, using this pulse profile and the corresponding distribution function as initial conditions 
for the Vlasov code, together with periodic boundary conditions, the temporal 
evolution of the Vlasov-Poisson system is monitored (see Sec.\ref{SubSec_Equations_Vlasov}).
A benchmark is presented 
to test the code versus known results (see Sec.\ref{SubSec_Results_bechmark}).
Different features of the pulses are followed such as
velocity, height, width, energy contours 
and the shape in both spatial direction and phase space
to show either stability (see Sec.\ref{SubSec_Results_Stability}) 
or instability (see Sec.\ref{SubSec_Results_Instability}) for different combinations of parameters. 
Conclusion remarks on the simulation results are presented in Sec.\ref{Sec_Conclusions}.


\section{Basic Equations and the Numerical Schemes} \label{Sec_Model}
Equations and variables are normalized 
to ionic parameters such as the ion plasma frequency, the ion Debye length 
and the ion thermal velocity.
The details are presented in table~\ref{table_normalization}. 
\begin{table}
\small
\caption{Normalization of quantities.}
\begin{ruledtabular}
\begin{tabular}{cccc}
   \multirow{2}{*}{Name}&   \multirow{2}{*}{Symbol}& 
   \multicolumn{2}{c}{Normalized by}  \\
  {}&  {}&  Name& formula \\
 \hline
  Time     & $\tau$  	&ion plasma frequency 		&$\omega_{pi}  = {\big(\frac{n_{i0} e^2}{m_i \varepsilon_0}\big)^{\frac{1}{2}} }$   \\
  Length   & $L$  	&ion Debye length		&$\lambda_{Di} = \sqrt{ \frac{\varepsilon_0 K_B T_i}{n_{i0} e^2}   }$   \\
  Velocity & $v$  	&ion thermal velocity		&$v_{th_i} = \sqrt{\frac{K_B T_i}{m_i}}$   \\
  Energy   & $\varepsilon$  	&ion thermal energy			&$K_B T_i$   \\
  Potential   & $\phi$  	&{-------}		&$\frac{K_B T_i}{e}$   \\
  Charge 		&$q$		&elementary charge &$e$ \\
  Mass 			&$m$		&ion mass		&$m_i$
\end{tabular}
\end{ruledtabular}
\label{table_normalization}
\end{table}

\subsection{Fully kinetic Sagdeev pseudo-potential approach} \label{SubSec_Equations_Sagdeev}
\subsubsection{Constructing Sagdeev pseudo-potential}
The Sagdeev approach starts with manipulation of Poisson's equation
\begin{equation*}
  \frac{ \dd{}^2 \phi(x,t)}{\dd x^2}  - \sum_s n_s = 0; \quad s=i,e
\end{equation*}
in which $n_s$ stands for the density of each species, namely electrons
($s = e$) and ions ($s = i$).
This equation is multiplied by $\big( \frac{\dd \phi}{\dd x} \big)$ and then integrated over $\phi$
to produce the Sagdeev pseudo-potential:
\begin{equation}
    S(\phi) = \int_0^{\phi} \sum_s n_s(\phi) \dd \phi - 1 + \textrm{const} \quad s=i,e \label{Eq_Sagdeev_potentail}
\end{equation}
After constructing $S(\phi)$, one can easily integrate once more over it and construct the 
electric potential $(\phi(x))$ and the electric field $(E(x))$.

In order to derive a solution, the species' densities should be computed as a function of 
electrical potential ($\phi$). 
Analytically, there are two approaches for such computation, i.e. fluid and kinetic.
In the first approach, fluid equations (including equation of momentum and continuity equation)
are used in a co-moving coordinates to find the density function. 
In the kinetic model, however, the density is achieved by direct integration
of distribution function over velocity:
\begin{equation}
    n_s(\phi) = n_{0s} \int f_s(\phi) \dd v. \label{Eq_density_calculation}
\end{equation}
in which, $n_{0s} = N_{0s} q_s$ (unperturbed number density multiplied by normalized charge) 
stands for unperturbed charge density. 
Due to the normalization used, unperturbed number density of each species
is equal to one $N_{0s} = 1$ and $q_e = -1, q_i = +1$.

Here, we have adopted the kinetic approach in our numerical model, and
hence a valid distribution function is needed as the starting point
of our numerical approach. 
We start by recalling a well-known property of the Vlasov equation; 
a distribution function that depends
in any way on the particle constants of motion is a solution of the Vlasov equation.
Specifically a distribution function that 
depends on the total energy of a particle 
(electrostatic energy $q\phi$ + kinetic energy) is a solution, 
provided we describe  
the system in a frame where $\phi$ is time-independent.
Since we in practice are interested in propagating 
(and thus time-dependent) solitary wave 
profiles, in the end we will need 
to transform solutions of this type into the laboratory frame.

In this study, we are focusing on two different 
types of distribution functions, namely the Maxwellian distribution function:
\begin{equation*}
    f(\varepsilon) = \sqrt{ \xi \frac{1}{2 \pi}}  \exp (- \xi \frac{\varepsilon}{2})
\end{equation*}
and the Kappa distribution function\cite{hellberg2009comment}:
\begin{equation*}
    f(\varepsilon) = 
    \frac{\Gamma (\kappa)}{\Gamma(\kappa- 1/2)}
    \sqrt{\frac{1}{2 \pi}\ \frac{\xi}{(\kappa - 3/2)}}
    \Big[ 1 + \frac{\xi}{(\kappa - 3/2)} \frac{v^2}{2}  \Big]    
\end{equation*}
in which $\xi = \frac{m}{m_i}\frac{T_i}{T}$ and $\Gamma$ stands for the usual gamma function.
The Kappa distribution function was proposed for the
first time by V. M. Vasyliunas \cite{vasyliunas1968survey}
when studying ``low-energy electrons in the evening sector of the magnetosphere''
to explain the long tail appearing in the velocity distribution at high (velocity) values.
It is designed to address the case of superthermal
particle populations.
This type of distribution function 
has been used for modeling
non-Maxwellian backgrounds in various astrophysical
and experimental plasma situations
\cite{pierrard2010kappa,sarri2010observation,maksimovic1997ulysses,hellberg2002generalized}.
The Kappa distribution function depends   
on a real parameter ($\kappa$), 
which measures the magnitude of the superthermal tail in the distribution function,
representing  the excess of highly energetic particles.
As $\kappa$ increases, the superthermal tail shrinks, and the Maxwellian distribution 
can be recovered at $\kappa \longrightarrow \infty$.

\subsubsection{Constructing the potential dependency}\label{SubSec_Equations_Schamel}
The aim of this subsection is to present a simple and easy way 
to derive the \emph{Schamel distribution function} from the 
energy point of view\cite{schamel_1}.
This approach can be easily implemented into the 
Vlasov-solver.
Furthermore it can be easily extend to different types
of distribution function 
(here we will consider Kappa distribution function).

Assuming a potential pulse moving with a velocity ($v_{p}$) in the laboratory frame, 
the following steps should be taken in order to consider its effect on the distribution fucntion:
\begin{enumerate}
    \item transforming the  kinetic energy into the co-moving frame:
    $\varepsilon'_k = \frac{1}{2} m v'^2$ in which $v' = v - v_{p}$.
    \item  finding the shifted velocity in the co-moving frame: 
    $v'_{sh}  = \textrm{sign} (v') 
    \sqrt{\frac{2|(\varepsilon'_{k} - \varepsilon_{\phi})|}{m}}$.
    \item transforming the kinetic energy back to the laboratory frame:
        $\varepsilon_{k_{sh}} =  \frac{1}{2} m v_{sh}^2  $ in which 
    $v_{sh} = v'_{sh} + v_{p}$. 
\end{enumerate}

The shifted velocity equation
indicates that the particle with positive potential energy 
($\varepsilon_{\phi} >0$)
has lost some portion of its kinetic energy due to 
the interaction with the potential 
and stored it as potential energy (in the co-moving frame).
Simply put, these particles see the potential as a ``hill''.
If the the kinetic energy of such particles are smaller than the 
maximum potential energy, they will be reflected by the potential,
hence called \emph{reflected particles}.
While particles possessing negative values 
for the potential energy ($\varepsilon_{\phi} < 0 $)
feels an acceleration in the co-moving frame
due to the push originating from the potential. 
In other words, they experience the potential as a hole 
and some of them  will be trapped inside it, hence called \emph{trapped particles}.

In order to add the effect of trapped particles, 
one can use a shifted Maxwellian distribution function for modeling them, 
as suggested by Schamel\cite{schamel_1}.
The trapped population can be achieved by following the steps below:
\begin{enumerate}
    \item the velocity is transformed into the moving frame: 
    $v' = v - v_{p}$.
    \item the shifted velocity
  $v'_{sh}  = \textrm{sign} (v') 
    \sqrt{\frac{2|(\varepsilon'_{k} - \varepsilon_{\phi})|}{m}}$
  and the shifted kinetic energy in the moving frame
  $\varepsilon'_{k_{sh}} = \frac{1}{2}m v'^2_{sh}$
  are calculated.    
    \item kinetic energy of the moving frame 
    ($\varepsilon_{p} = \frac{1}{2} m v_{p}^2$) is added to 
      the kinetic energy of the particles in the moving frame ($\varepsilon'_{k_{sh}}$) 
      with a coefficient $\beta$: 
      $ \varepsilon_{t} = \varepsilon_{p} + \beta \varepsilon'_{k_{sh}}$.
\end{enumerate}

Based on $\beta$, the distribution function of trapped particles 
can take three different types of shapes, 
namely \emph{hole} ($\beta<0$), \emph{plateau} ($\beta = 0$) and \emph{hump} ($\beta>0$).
Note that since the above method is implemented for both positive and negative 
potential energy, the two cases of  reflected and trapped particles are considered 
in the model self-consistently.

Hence, the total form of the distribution function can be written as follows:
\begin{equation}
  f (\varepsilon)=  
 \begin{cases} 
      f(\varepsilon_f)  \quad |v'| >  \sqrt{\frac{2\varepsilon_{\phi} }{m}} \\ 
      f(\varepsilon_t) \quad |v'| <  \sqrt{\frac{2\varepsilon_{\phi} }{m}}
  \end{cases} 
  \label{Schamel_Dif_energy}
\end{equation}
in which:
\begin{align*}
    \varepsilon_{f} &=  \varepsilon_{k_{sh}} \\
    \varepsilon_{t} &= \varepsilon_{p} + \beta \varepsilon'_{k_{sh}}. 
\end{align*}
The above distribution function (written in form of 
kinetic energy)
can be shown to be equivalent to
the Schamel distribution function\cite{schamel_1}.

\subsubsection{Summary}
In summary, the path to find the Sagdeev solutions $(\phi(x))$ consists 
of four steps:
\begin{enumerate}
   \item $f(\phi)$: by using the above recipe (Sec.\ref{SubSec_Equations_Schamel}), the distribution function's dependency 
on the potential is calculated (Eq.~\ref{Schamel_Dif_energy})
    \item $n(\phi)$: by integration over velocity, the density can be obtained (Eq.~\ref{Eq_density_calculation}).
    \item  $S(\phi)$: by following the same routine for all the species, 
the Sagdeev pseudo-potential can be constructed by integration 
over $\phi$ (Eq.~\ref{Eq_Sagdeev_potentail}).
    \item $\phi(x)$: the potential profile $\phi(x)$ can found 
    from the Sagdeev pseudo-potential $S(\phi)$ by integrating (Eq.~\ref{Eq_Sagdeev}).
\end{enumerate}
Three variables affect the Sagdeev pseudo-potential ($S(\phi)$) and pulse potential profile ($\phi(x)$) ,
namely the shape of the distribution function, the velocity of the solitary wave ($v_{p}$)
and the trapping parameter ($\beta$).

While the above procedure does generate steady state solutions, 
there is no guarantee these solutions are stable.  
One should note that through out the derivation of the Sagdeev solutions, 
there is no temporal evolution in the model.
In the Sec.\ref{SubSec_Results_Stability},
we will study the long-time behavior (i.e. stability )  of Sagdeev solutions,
employing fully kinetic Vlasov-Poisson method. 


\subsection{Fully kinetic Vlasov-Poisson method} \label{SubSec_Equations_Vlasov}
Here, the study is restricted to electron-ion plasmas, 
and both species' dynamics are followed by the Vlasov equations:
\begin{multline}
\frac{\partial f_s(x,v,t)}{\partial t} 
+ v \frac{\partial f_s(x,v,t)}{\partial x} 
\\ +  \frac{q_s}{m_s} E(x,t) \frac{\partial f_s(x,v,t)}{\partial v} 
= 0, \ \ \  s = i,e
\label{Eq_Vlasov}
\end{multline}
while Poisson's equation provides the force (electric) field:
\begin{equation}
\frac{\partial^2 \phi(x,t)}{\partial x^2}  = n_e(x,t) - n_i(x,t)
\label{Eq_Poisson}
\end{equation}
where $s = i,e$ represents the corresponding species.
Vlasov and Poisson equations 
are coupled by density integrations 
for each species to form a closed set of equations:
\begin{equation*}
  n_s(x,t) = n_{0s} N_s(x,t), 
\end{equation*}
\begin{equation}
N_s(x,t) = \int f_s(x,v,t) dv.
\label{Eq_number_density}  
\end{equation}
In which $N_s$ stands for the number density.
$n_{0s}(=N_{0s}q_s)$ is the (normalized) unperturbed value of the charge density.
The quasi-neutrality condition stays true:
\begin{equation*}
\sum_s n_{0s} = \sum_s N_{0s} q_s = 0.
\end{equation*}

The kinetic simulation approach utilized here 
is based on the Vlasov-Hybrid Simulation (VHS) method in which 
a distribution function is modeled by phase points
\cite{nunn1993novel,kazeminezhad2003vlasov,jenab2011preventing}. 
The arrangement of phase points in the phase space 
at each time step provides the distribution function,
and hence all the kinetic momentums 
such as density, entropy, etc. 
The initial value of distribution function associated to each of
the phase points stay intact
during simulation.
This advantage guarantees the positiveness of distribution function
under any circumstances\cite{kazeminezhad2003vlasov}.
VHS method can be interpreted as a
PIC code of uniform weighting with lower noise level.

Each temporal update of the simulation consists of three steps 
which are summarized below.
\begin{enumerate}
    \item Integration of the distribution function over velocity direction 
    to achieve the number density (Eq.~\ref{Eq_number_density}).
    \item Calculation of the electric potential and field 
    by solving Poisson's equation (Eq.~\ref{Eq_Poisson}).
    \item Determination of the new arrangement of 
    phase points in the phase space for the next step
by solving the Vlasov equation for each species based on 
the characteristics method utilizing a leap-frog scheme (Eq.~\ref{Eq_Vlasov}).
\end{enumerate}

As for the initial condition of our Vlasov simulations, 
we need to have the distribution function of each of the species. 
In order to implement the calculated pulse profile $(\phi(x))$ self-consistently,
we use the distribution function described in Sec.\ref{SubSec_Equations_Schamel}
to produce the initial distribution functions, 
which have the trapped and reflected population inside them.
Furthermore, the initial conditions include some parameters and variables
which need to set before the simulation starts. 
These are explained in details in the next section.

\subsection{variables and parameters} \label{SubSec_Equations_variables}
The constant parameters which remain fixed through all
of our simulations include: 
the mass ratio $\frac{m_i}{m_e} = 1836$,
the temperature ratio $\frac{T_e}{T_i} = 20$
and  $\Delta x = 0.25$ 
where $\Delta x$ is the grid size on the spatial direction.
The length of the simulation box is  $L = 256$ or $L = 2048$, 
which is specified in each section.
Periodic boundary conditions are adopted on the spatial direction. 
Furthermore,
the input parameters for each set of the simulation set consists of three
variables:
\begin{itemize}
\setlength\itemsep{0.0em}
    \item the shape of the distribution function, either Maxwellian or Kappa
    \item the trapping parameter ($\beta$)
    \item the velocity of the pulse ($v_p$)
\end{itemize}

Note that by ionic normalization (table \ref{table_normalization}),
the ion sound velocity, the electron plasma frequency
and the electron thermal velocity are 
$v_C = \sqrt{1 + \frac{T_e}{T_i}} = 4.5$,
$\omega_{pe} = \sqrt{\frac{m_i}{m_e}} = 42.8$
and $v_{th_{e}} = \sqrt{\frac{m_i}{m_e}\frac{T_e}{T_i}} = 191.6$;
respectively.
In order to follow the form of many theoretical papers. 
In what follows, we will express the velocity of solitary waves 
using the \emph{Mach number}: $M = \frac{v_{p}}{v_C}$.

In the case of $L=256$ the simulation grid consists of 
$(S_x, S_v) = (1024,3000)$ cells for electrons phase space and 
$(S_x, S_v) = (1024,8000)$ for ions. 
The difference between the species phase space is in the velocity direction,
the electron velocity cut-off is $(v_{min}, v_{max}) = (-1000,+1000)$
while for ions it is $(v_{min}, v_{max}) = (-6,+14)$. 
Initially there are 16 phase points per cell in the phase space. 
Hence the total number of phase points for electrons (ions) 
is $\Sigma_e=65\times10^6$ ($\Sigma_i = 131\times10^6)$.
For the case of $L=256$, grid size is $S_x = 4096$ 
and hence $\Sigma_e=260\times10^6$ and $\Sigma_i = 524\times10^6$.

Since the dynamical part of the simulation method
utilizes the Vlasov equation, which is the collision-less Boltzman equation, 
simulations must conserve entropy and other forms of Casimir invariants \cite{elskens2014vlasov} as well as the 
total energy. 
It is vital for any (collision-less) kinetic simulation to 
keep track of these conservation laws, as they provide 
extremely valuable indicators about the validity of the simulation results.
They are the guardians of simulation codes against bugs in either algorithm or 
implementation. 
Therefore, we have presented the temporal evolution 
of the deviation (called from here on as ``deviation in percentage''):
\begin{equation*}
  D_x = \frac{X(\tau) - X(0)}{X(0)} \times 100\%
\end{equation*}
for all the simulations reported in this study
in order to be fully transparent about the research presented here.
$X$ represents the total energy ($\varepsilon_t = \varepsilon_k + \varepsilon_{\phi}$) and 
entropy ($P_s = \int \int f_s \ln f_s \dd v \dd x $).
$\varepsilon_k = \int \int \frac{1}{2} m v^2 f_s \dd v \dd x$ 
implies kinetic energy 
and $\varepsilon_{\phi} = \int \int q \phi_s \dd v \dd x$ indicates potential energy of the system.


\section{Results and Discussion} \label{Sec_Results}
 
  \subsection{Benchmark: the effect of ion dynamics on standing electron holes} \label{SubSec_Results_bechmark}
  \begin{figure*}
   \subfloat{\includegraphics[width=0.41\textwidth]{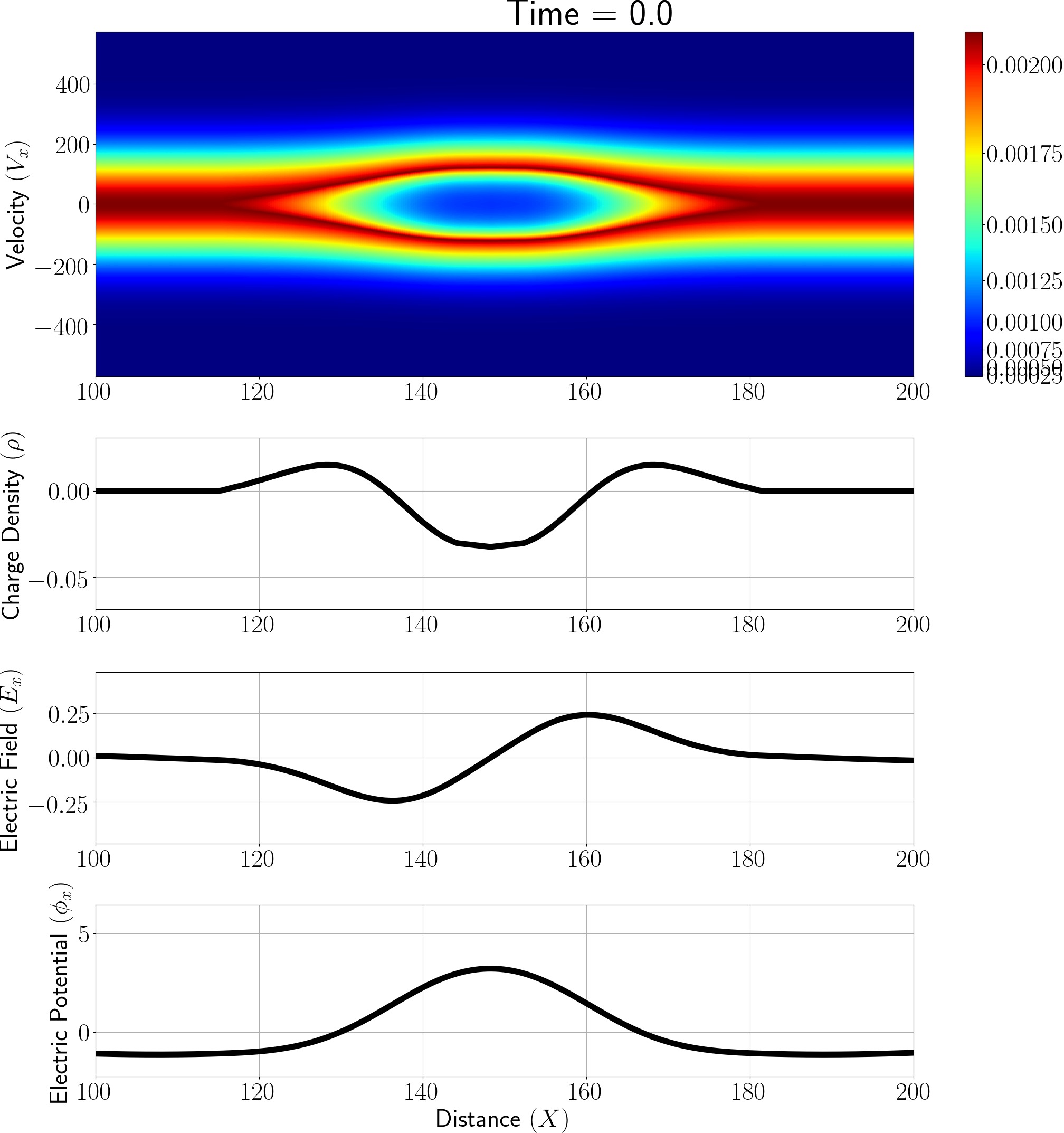}} \hspace{1.0cm}
   \subfloat{\includegraphics[width=0.41\textwidth]{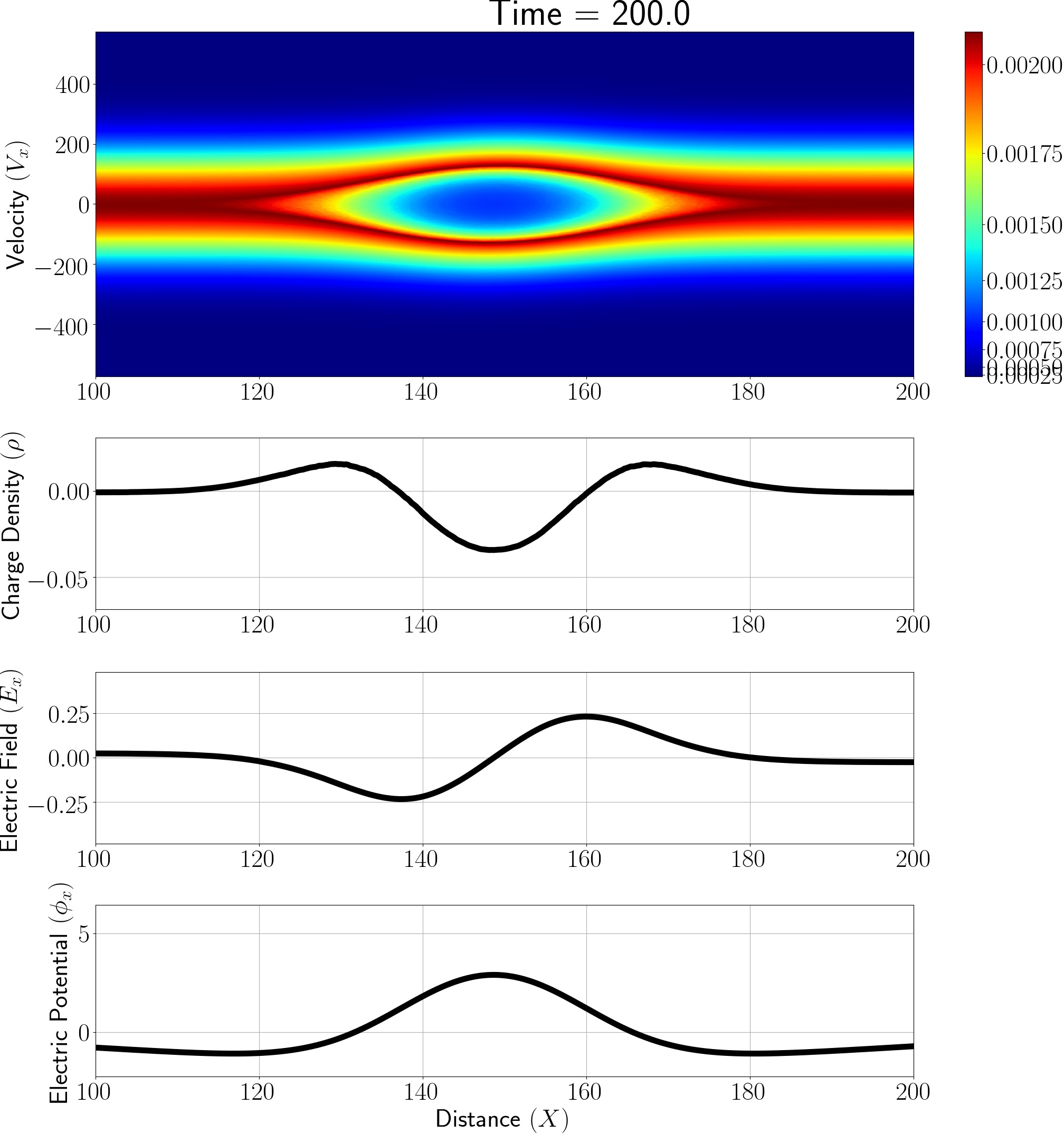}} \\ 
   \subfloat{\includegraphics[width=0.9\textwidth]{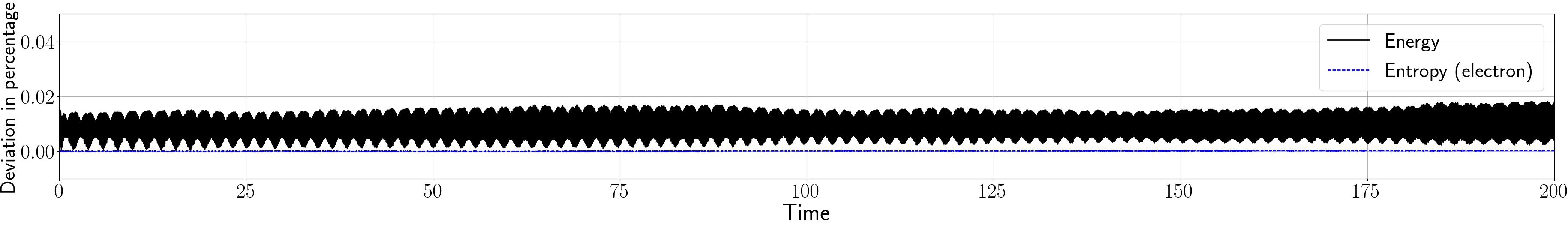}}
   \caption{Simulation results for $M=0$, $\beta = -3.5$ when ions are motionless is presented in the lab frame. 
   Temporal evolution of electron distribution function (first row), charge density (second row), 
   electric field (third row) and electric potential (fourth row) is shown for two different 
   time steps, e.g. $\tau = 0, 200$.
   Temporal evolution of deviation from the initial value is also depicted 
   for energy and entropy (of electrons) in the last row.
   It indicates that energy deviation stays below $0.02\%$ and 
   as for electron entropy, deviation is almost zero. 
   Electron holes stand still when ions are considered as fixed species in the back ground
  providing the positive charge.}
  \label{Fig_M0_B-3o5_Max_e_without_ions}
\end{figure*}
\begin{figure*}
   \subfloat{\includegraphics[width=0.33\textwidth]{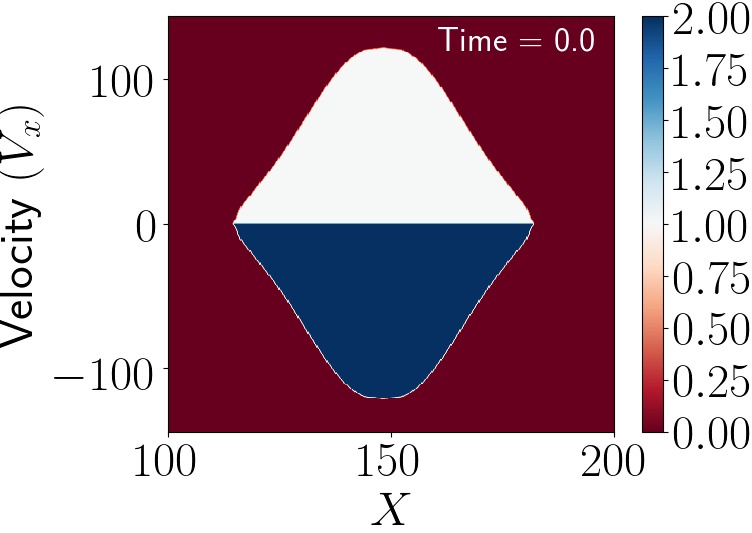}} 
   \subfloat{\includegraphics[width=0.33\textwidth]{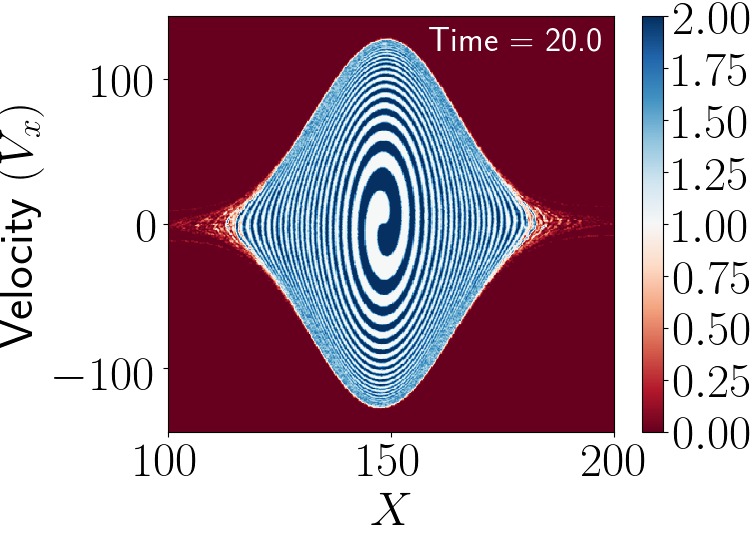}} 
   \subfloat{\includegraphics[width=0.33\textwidth]{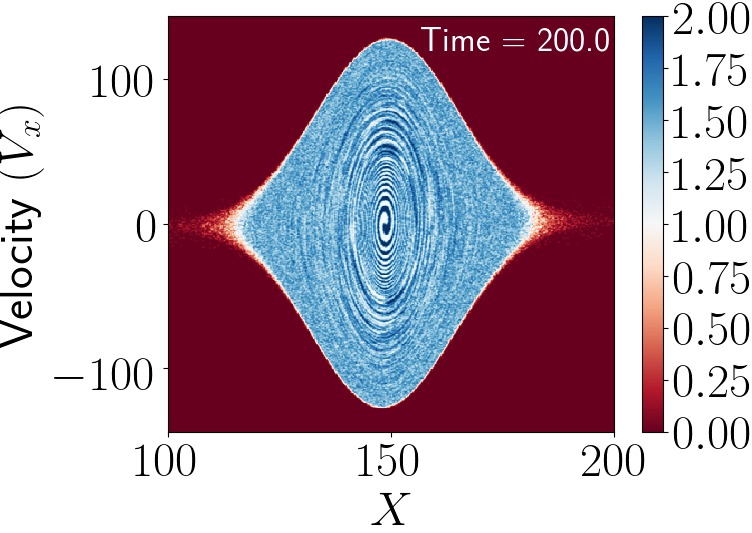}}
   \caption{ The dynamics inside an standing electron hole is shown 
   for the case of Fig.~\ref{Fig_M0_B-3o5_Max_e_without_ions} in the lab frame.
   In the initial condition, the upper/lower part of the electron hole is marked with 
   color white/blue (with marker value equal to $1$ or $2$) in three time steps ($\tau = 0, 20, 200$). 
   As temporal evolution progresses these two colors intertwine each other
   reflecting the trajectories of the phase points in the phase space. 
   Note that during the long-time evolution of electron hole, trapped particles
   stay in the boundaries of trapped region which shows the high-precision of the simulation code.}
  \label{Fig_MPP_M0_B-3o5_Max_e_without_ions}
\end{figure*}

In this section, the well-known example\cite{saeki1998electron,saeki1991stationary} of the effect of ion dynamics on an initially 
standing electron hole is taken as a benchmark of our simulation code. 
It is considered a well-established fact that an electron hole with zero velocity
will stand fixed if the ions dynamics is removed from the theory. 
This is shown in both simulations\cite{saeki1998electron} 
and theoretical studies\cite{saeki1991stationary}.
However, when the ion dynamics is considered 
(the ions can move and react to the field)
the result is totally different. 
In the electron phase space, 
the standing hole breaks up into two oppositely moving holes,
which are completely symmetric to one another since 
there is no disparity between negative and positive velocities.
The initial profile of the electron hole is achieved 
using the Sagdeev approach with inputs $M = 0.0, \beta = -3.5$
by considering electron following the Maxwellian distribution function,
and ions as motionless species providing the background positive charge.

\begin{figure*}
  \subfloat{\includegraphics[width=0.4\textwidth]{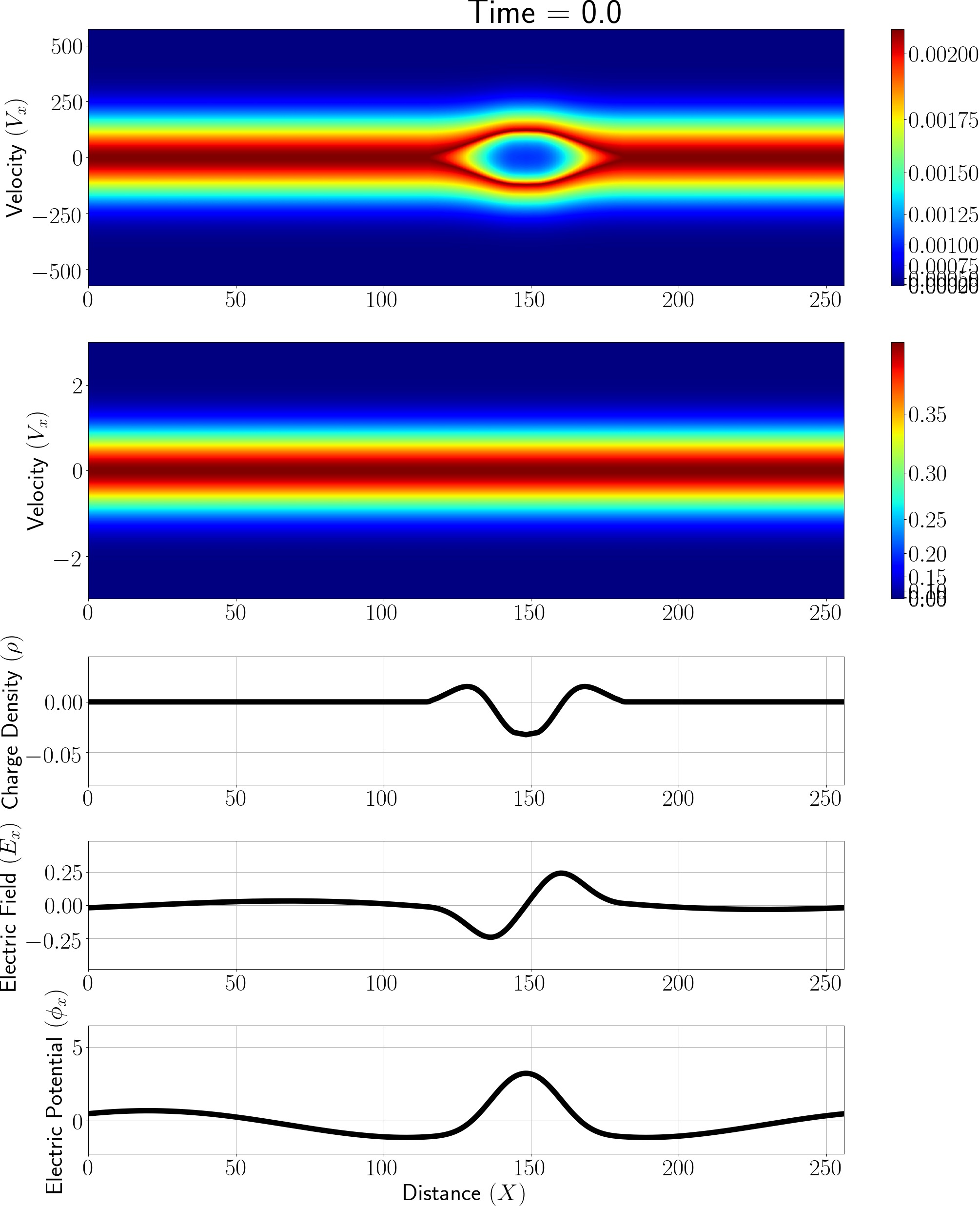}} \hspace{1.0cm}
  \subfloat{\includegraphics[width=0.4\textwidth]{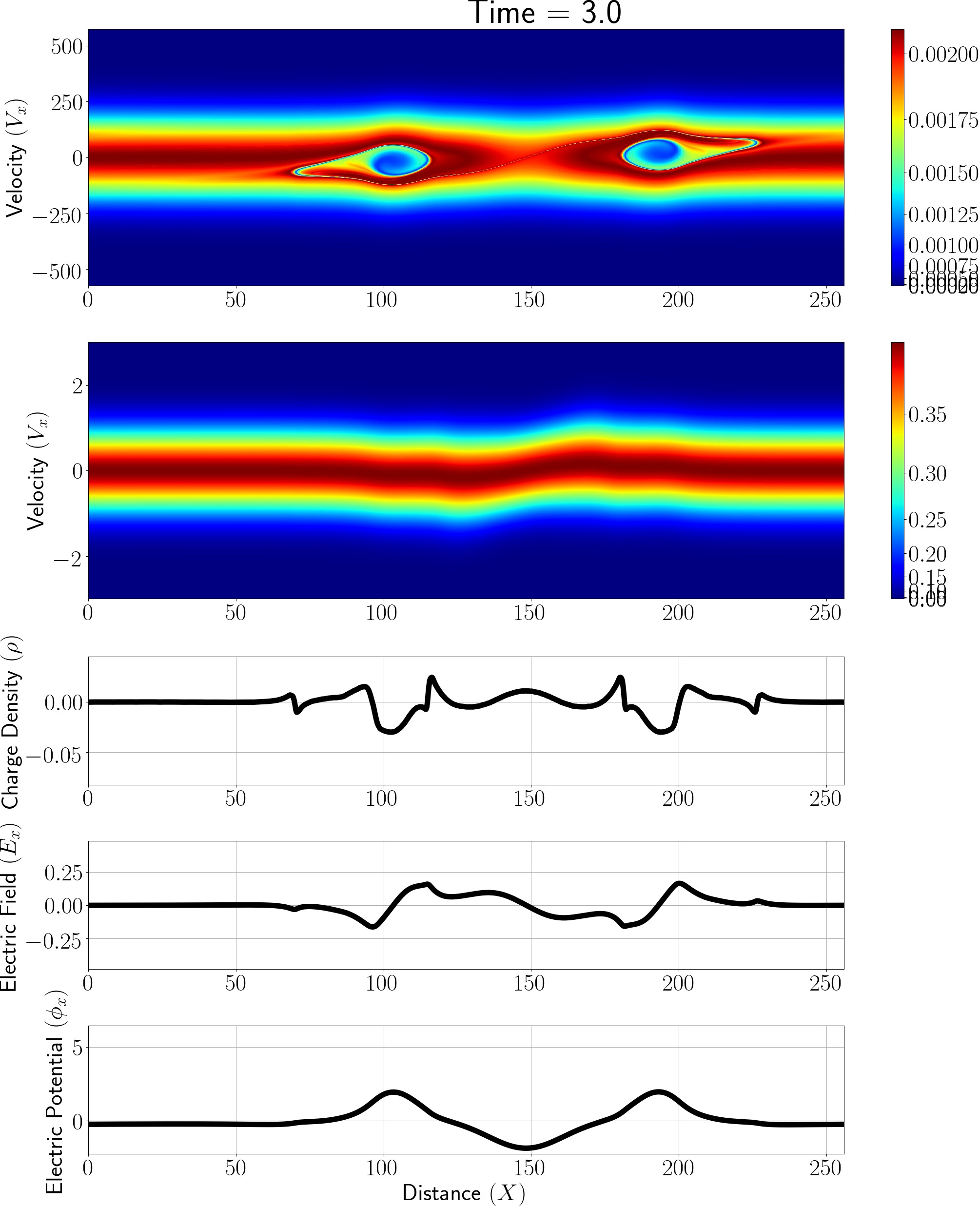}}\\
  \subfloat{\includegraphics[width=0.9\textwidth]{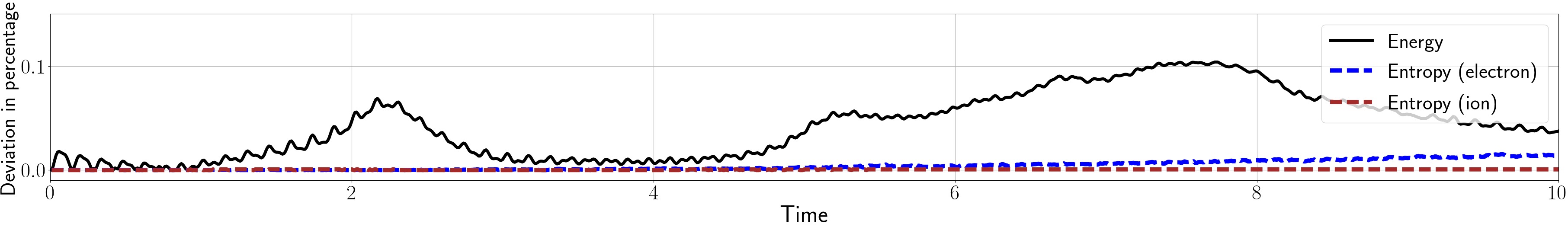}}
  \caption{Simulation results for $M=0$, $\beta = -3.5$ for an electron-ion plasma is presented in the lab frame.
	   The temporal evolution of quantities in both phase and physical space 
	   is shown for two different time steps, e.g. $\tau = 0, 3$.
	   Quantities are depicted in different rows, starting from the top row the include:
	   electron distribution function, ion distribution function, charge density, 
	   electric field and electric potential, deviation from conservation laws.
	   Electron hole breaks up when ions dynamics 
	   is considered in the simulation in contrast to simulations 
	   without ions dynamics (see Fig.~\ref{Fig_M0_B-3o5_Max_e_without_ions}).
	   Note the symmetry between left and right propagating electron holes 
	   after the breakup in distribution functions and charge density profiles.
	   Conservation of energy and entropy (for both electrons and ions) 
	   are sketched which shows deviation stays below $\%0.1$.}
  \label{Fig_M0_B-1_with_ions}
\end{figure*}

Fig.~\ref{Fig_M0_B-3o5_Max_e_without_ions} presents the results of a simulation 
without the dynamics of ions. 
The results show that the electron hole will stand still as 
predicted.
When the dynamics of ions is added to the simulation 
(see Fig.~\ref{Fig_M0_B-1_with_ions}),
the electron hole breaks up into two oppositely propagating holes. 
The break-up happens in the following steps. 
First, the positive potential of the electron hole compresses the ions 
on each side of the hole. 
Each of ion pulses creates its own positive potential, 
pulling the trapped electron population from both sides. 
This is felt by the electron hole as a
drag force and causes it to 
elongate and finally breaks up into two \cite{saeki1998electron}.

Three things should be noted which can establish the 
correctness and precision  of the simulation approach in these
two test simulations. 
In case of the first simulation,
the trapped population has to be confined in the trapped region 
for the entire time of the simulation. 
Fig.~\ref{Fig_MPP_M0_B-3o5_Max_e_without_ions} shows that 
despite the rapid dynamics inside the trapped population, 
no considerable leakage of the trapped population can be observed. 
This verifies the two parts of our numerical approach, i.e. Sagdeev and
Vlasov-Poisson codes. 
This shows that Sagdeev code is able to deliver the steady-state solution properly.
It also implies that the Vlasov-Poisson code can follow
the dynamics on the nonlinear stage free from any
numerical anomaly.

In case of the second simulation, 
the symmetry of the dynamics in the positive and negative side of 
phase space should be observed. 
The simulation code can follow the symmetry between the 
two moving holes with high precision indicating that the 
numerical code is capable of modeling nonlinear processes. 
Furthermore, the conservation of energy and entropy should be valid 
for the entire simulation time. 
In both cases, the deviation of these variables stay
below $0.1\%$.

  \subsection{Stability of the Sagdeev solutions} \label{SubSec_Results_Stability}

Initially, we are presenting the results of simulations carried out employing 
the Maxwellian distribution functions for both electrons and ions.
The first simulation (case I) 
includes the electron hole with $M = 1.28$ and $\beta = 0.0$. 
The flat-topped area in the electron distribution function indicates the value
$\beta = 0.0$. 
Long running simulation ($\tau \leq 200$) verifies the stability of the solution. 
The temporal evolution of the system  in presented in three kinds of plots to 
analyze this stability.

\begin{figure*}
  \subfloat{\includegraphics[width=0.4\textwidth]{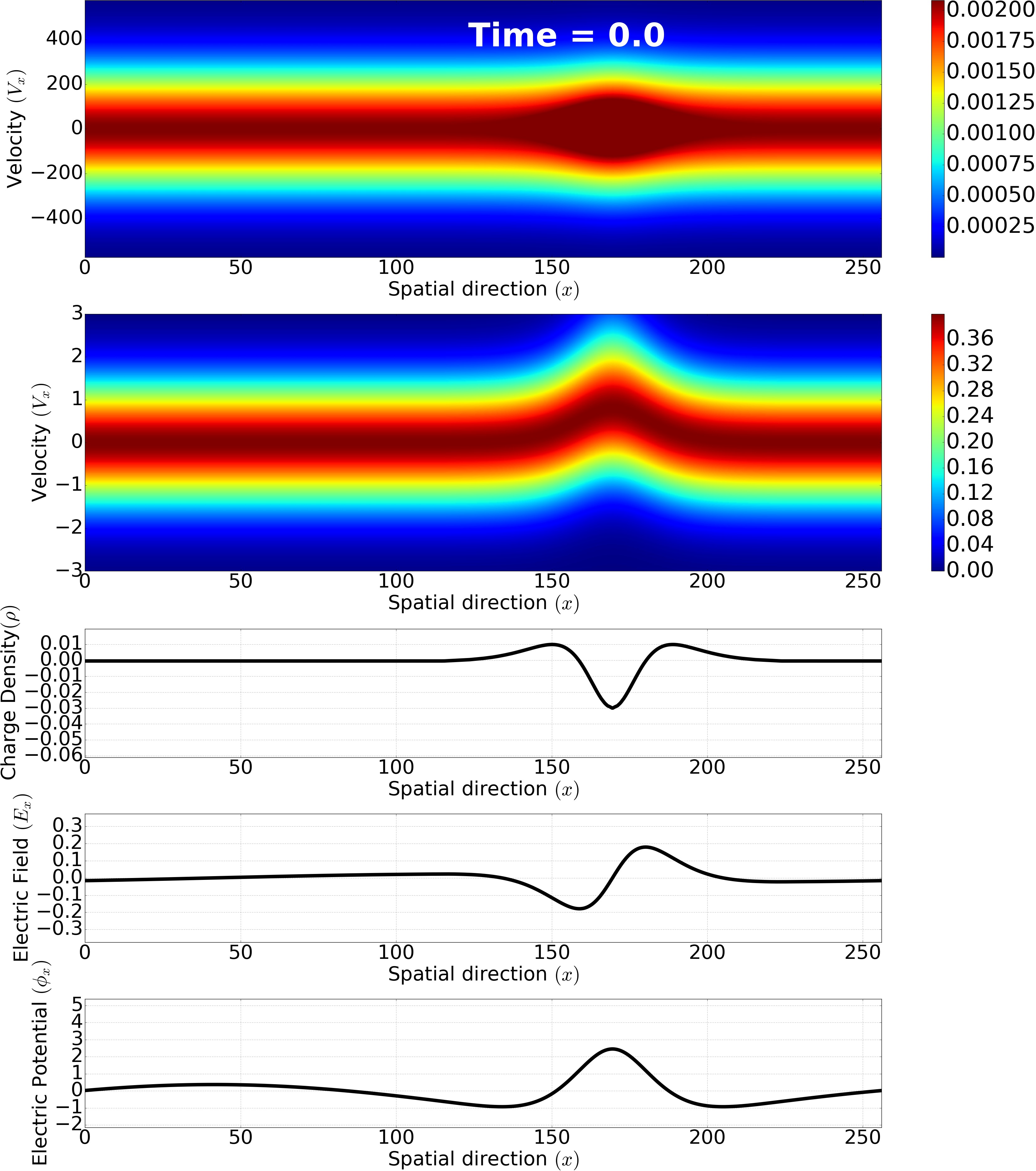}} \hspace{1.0cm}
  \subfloat{\includegraphics[width=0.4\textwidth]{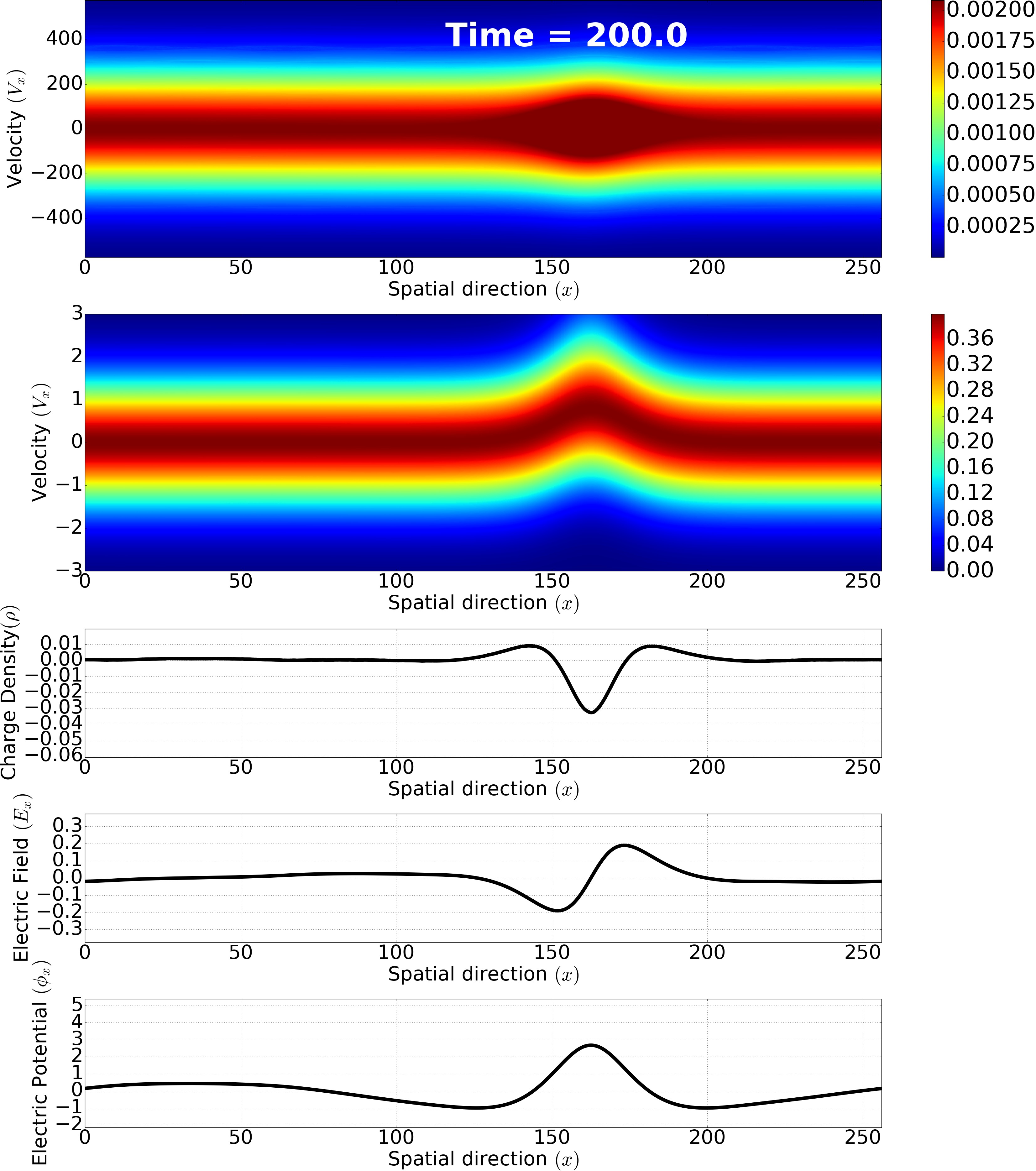}} \\
  \subfloat{\includegraphics[width=0.9\textwidth]{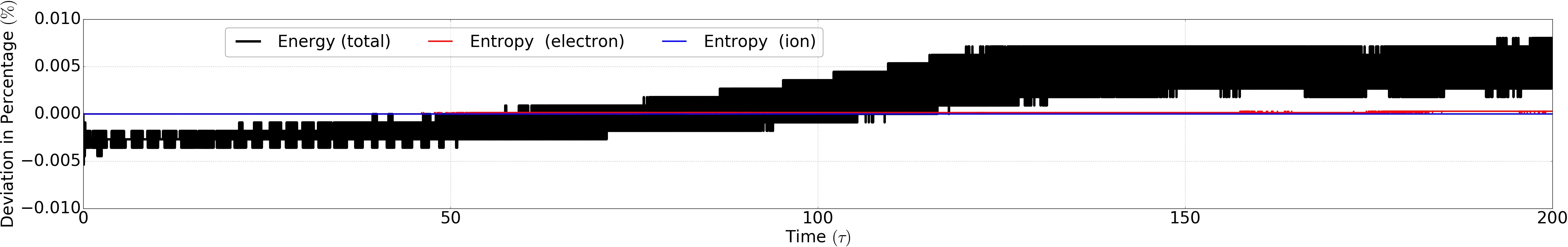}}
  \caption{Simulation results for $M = 1.28$ and $\beta = 0$ 
	   when distribution functions of both species are Maxwellian is presented in the frame $x-v_s \tau$. 
	   Temporal evolution of electron distribution function (first row),
	   ion distribution function (second row), charge density (third row), 
	   electric field (fourth row) and electric potential (fifth row)
	   is shown for first and last time step, e.g. $\tau = 0, 200$. 
	   Conservation of energy and entropy is presented for the simulation which
	   shows deviation below $\%0.1$. 
	   The stability of the Sagdeev solution for a long-time propagation can be observed. }
  \label{Fig_M1o28_B0_Max_e_Max_i}
\end{figure*}

Firstly, profiles of physical quantities in both phase space and real space 
such as distribution function, charge density, electric potential and field 
are depicted in Fig.~\ref{Fig_M1o28_B0_Max_e_Max_i}. 
It shows the state of the system in the beginning of the simulation (initial condition) and
at the last time step ($\tau = 200.0$). 
Since the time step is $dt = 0.01$, the last state in Fig.~\ref{Fig_M1o28_B0_Max_e_Max_i}
comes after $2\times10^4$ computational time steps. 
This long time simulation can easily show any instability or deformation 
in the profile of the solitary waves.
As can be observed, the profile stands unaltered during the simulation. 
Fig.~\ref{Fig_M1o28_B0_Max_e_Max_i} also displays the temporal evolution of energy and entropy, 
which shows robust conservation with deviation less than $0.01\%$.

\begin{figure*}
  \subfloat{\includegraphics[width=0.45\textwidth]{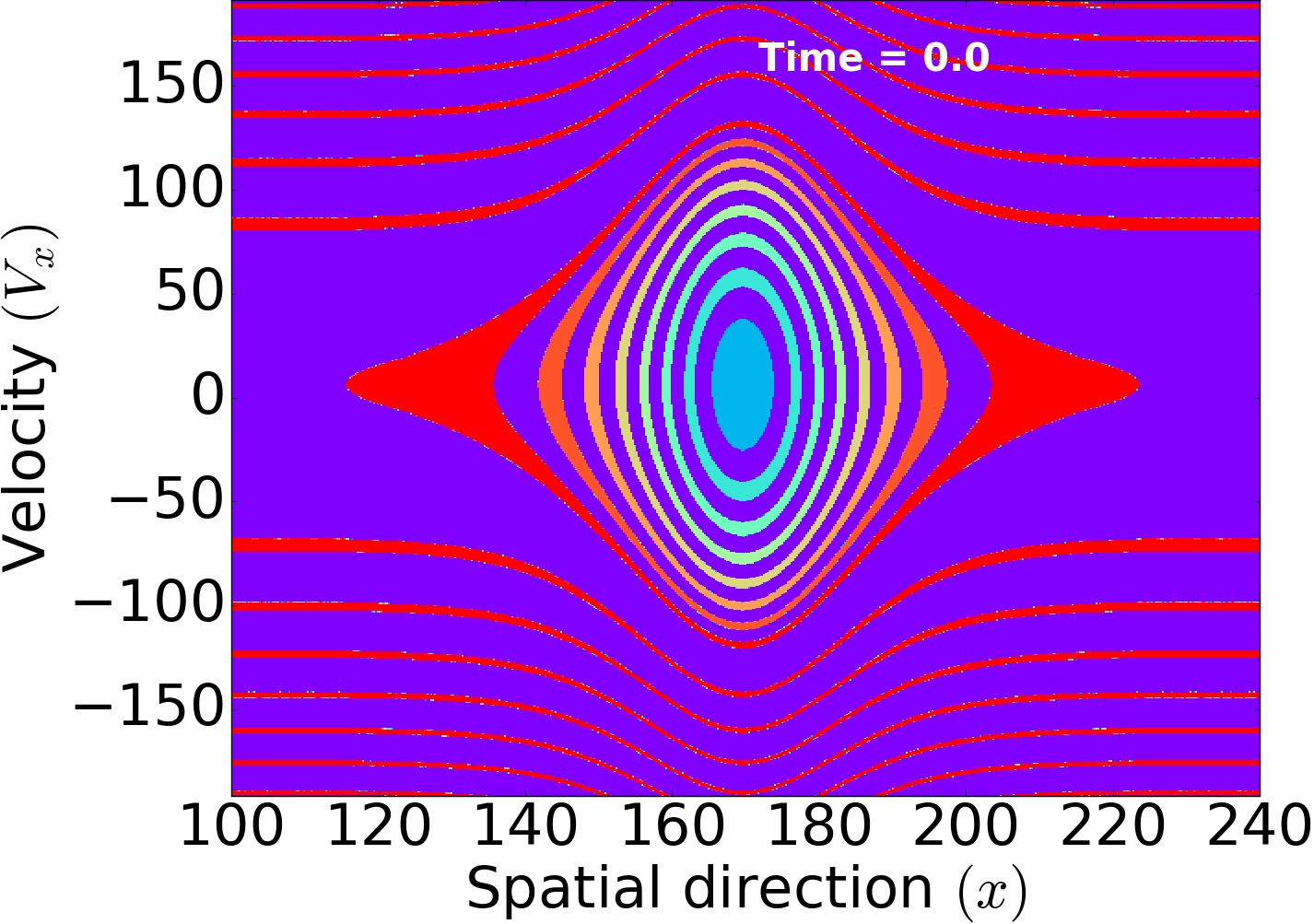}} \hspace{0.5cm}
  \subfloat{\includegraphics[width=0.45\textwidth]{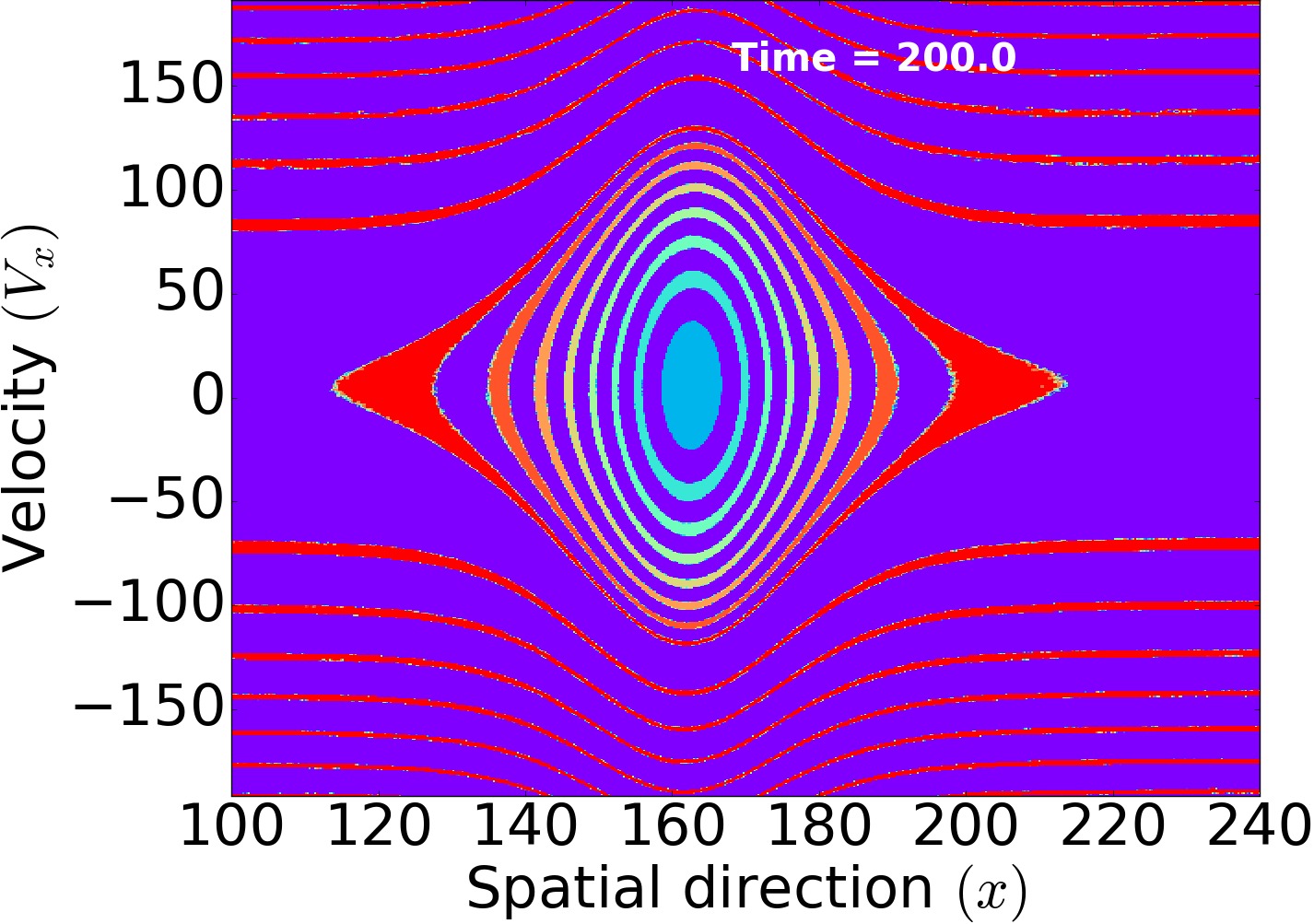}} 
  \caption{The energy contours in the phase space are presented for electron 
  distribution function for two time steps, e.g. $\tau = 0, 200$ of the case I ($M = 1.28$ and $\beta = 0$) in the frame $x-v_s \tau$.
  Trajectories of trapped particles can be recognized as circle loops.}
  \label{Fig_MPP_electron_T0_M1o28_B0_Max_e_Max_i}
\end{figure*}
\begin{figure*}
  \subfloat{\includegraphics[width=0.45\textwidth]{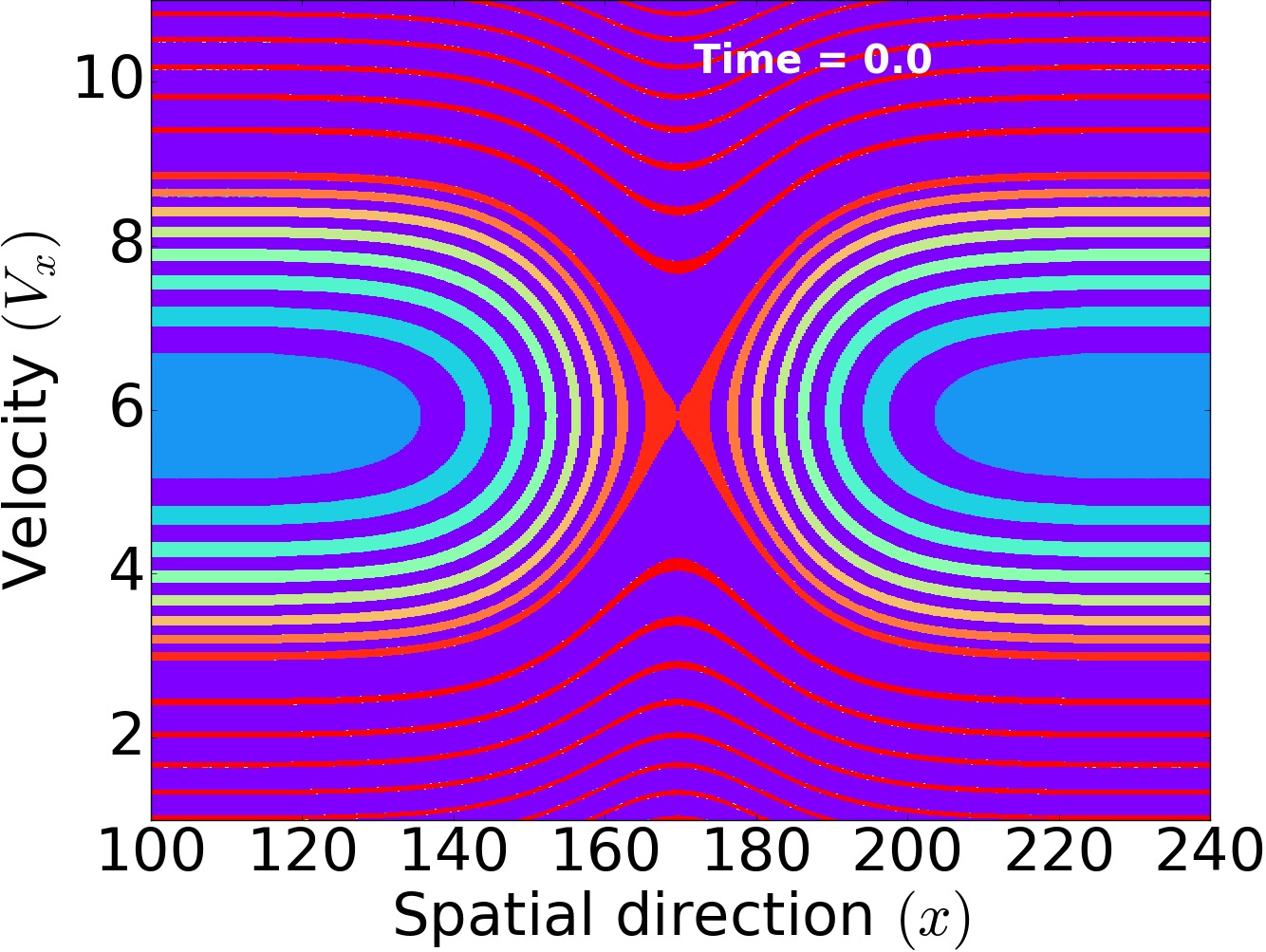}} \hspace{0.5cm}
  \subfloat{\includegraphics[width=0.45\textwidth]{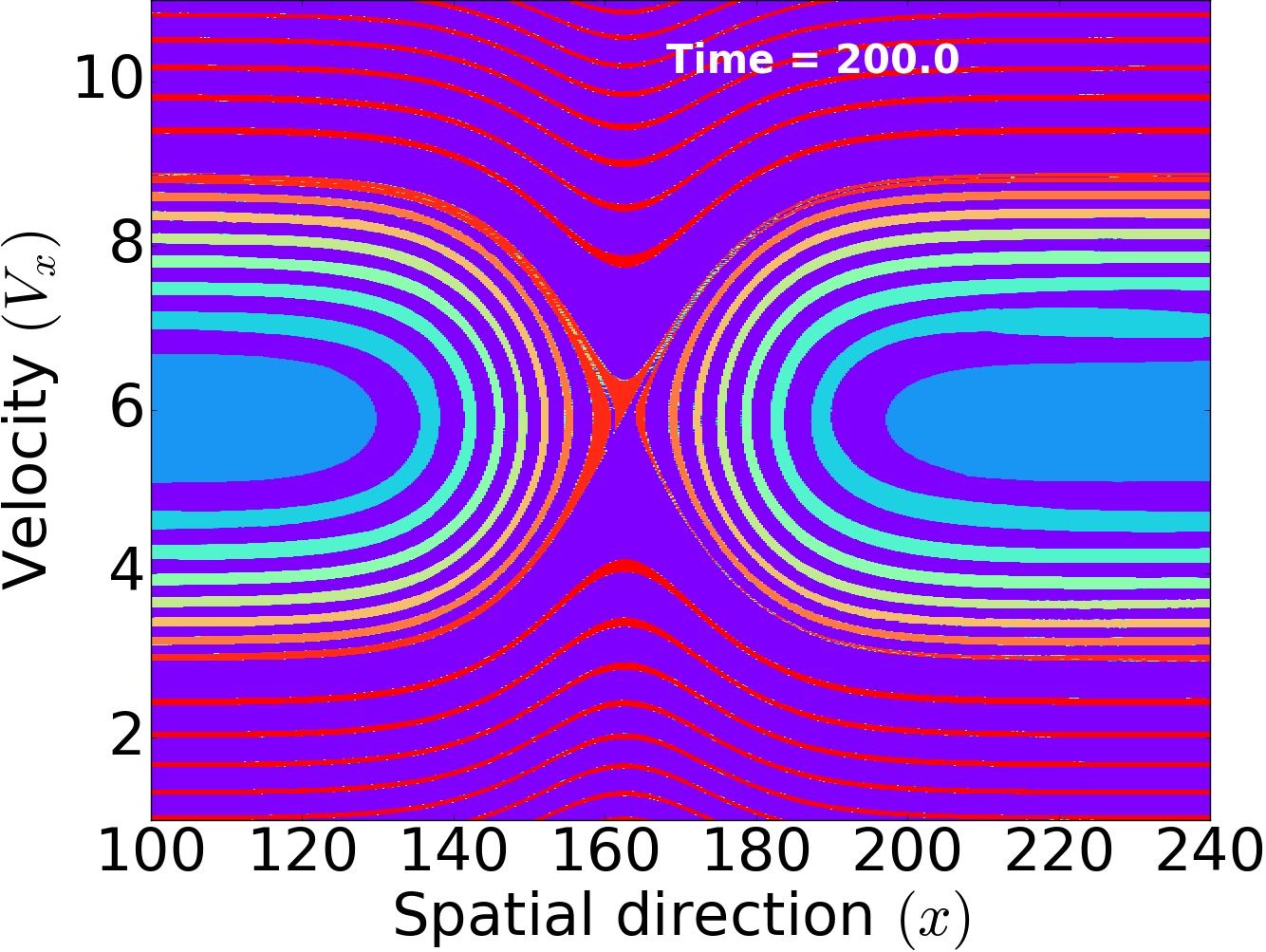}} 
  \caption{Temporal evolution of energy contours in the phase space is presented for ion 
  distribution function for two time steps, e.g. $\tau = 0, 200$ of the case I ($M = 1.28$ and $\beta = 0$) in the frame $x-v_s \tau$. 
  The energy trajectories of reflected particles can be seen in form of closed path versus free particles which have
  open trajectories.}
  \label{Fig_MPP_ion_T0_M1o28_B0_Max_e_Max_i}
\end{figure*}

In order to provide a clear-cut image of the phase space 
evolution, the initial phase points are marked with extra markers based
on their energy. 
These markers are just used for diagnosis purposes and has no
effect or involvement in the dynamical process. 
By depicting these markers in the phase space, it is possible 
to follow the evolution of the energy trajectories. 
Stability of the self-consistent solution demands that these 
trajectories should stay intact despite the rapid dynamics of the
phase space. 
Any unstable solutions 
can cause these trajectories 
to deform and change during the temporal evolution.
Fig.~\ref{Fig_MPP_electron_T0_M1o28_B0_Max_e_Max_i}
shows the stability of these trajectories 
for electron phase space in case I of the simulations.
The closed rings among these trajectories refers to the trapped electrons, 
which are marked with negative values, 
reflecting that their energy is smaller than 
the potential energy keeping them trapped.

The trajectories of energy are presented for ions as well,
which reflects the dynamics of reflected ions
(see Fig.~\ref{Fig_MPP_ion_T0_M1o28_B0_Max_e_Max_i}). 
Depicting the distribution function of these particles 
can not represent the dynamics. 
Since the value of the distribution function for this area of the phase space
is close to zero, at least ten orders of magnitude 
smaller than the maximum value of the distribution function around 
$v = 0$. 
Therefore, the reflected ions population are presented 
using the markers' values in stead of the distribution function.

\begin{figure*}
  \subfloat{\includegraphics[width=0.3\textwidth]{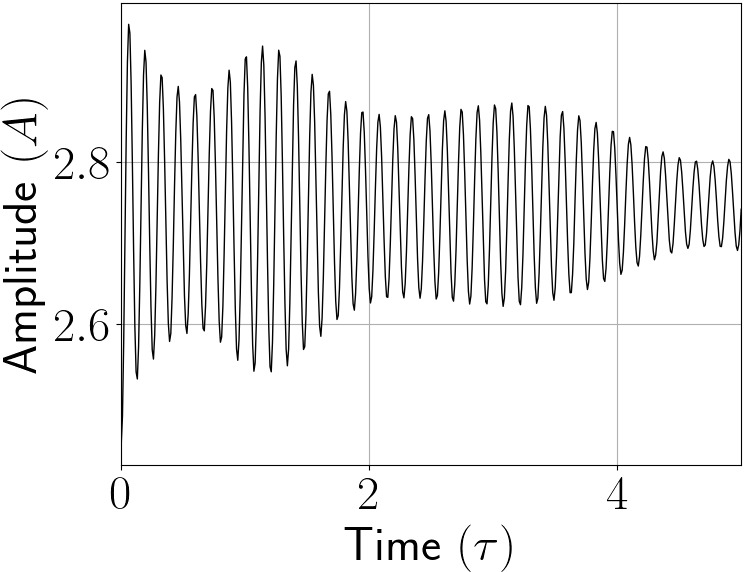}} \hspace{0.2cm}
  \subfloat{\includegraphics[width=0.3\textwidth]{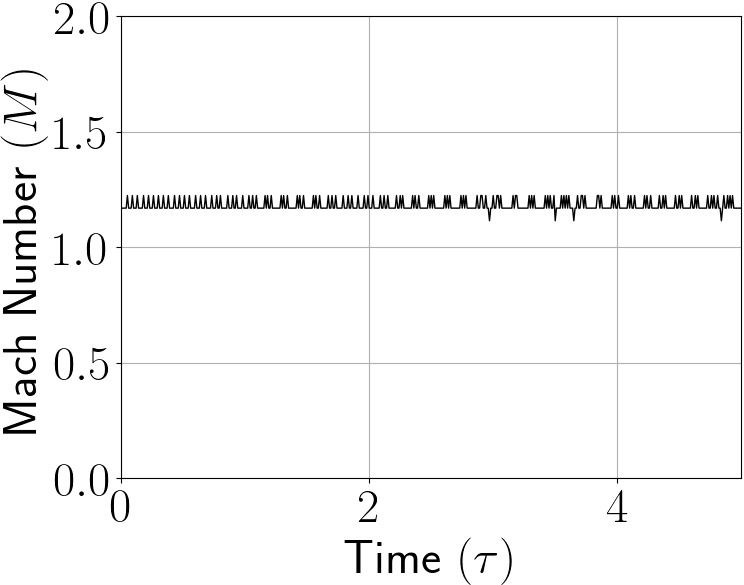}} \hspace{0.2cm}
  \subfloat{\includegraphics[width=0.3\textwidth]{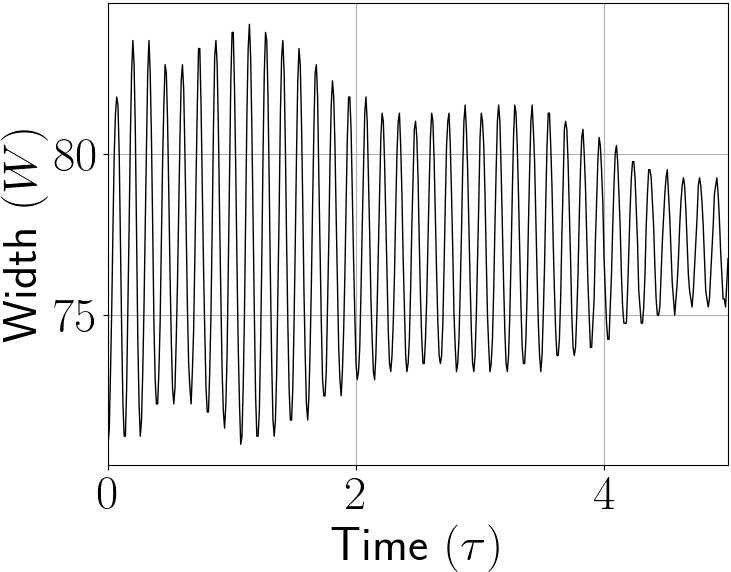}}
  \caption{Temporal evolution of three major features,
  e.g. amplitude, width and Mach number of ion-acoustic solitary waves are presented
  for short time ($\tau < 5.0$) for the case I ($M = 1.28, \beta = 0$). 
  The effect of Langmuir waves propagation can be observed as a rapid oscillation
  in amplitude and width of the ion-acoustic solitary waves.
  }
  \label{Fig_M1o28_B0_Max_e_Max_i_soliton_analysis_short}
\end{figure*}

\begin{figure*}
  \subfloat{\includegraphics[width=0.3\textwidth]{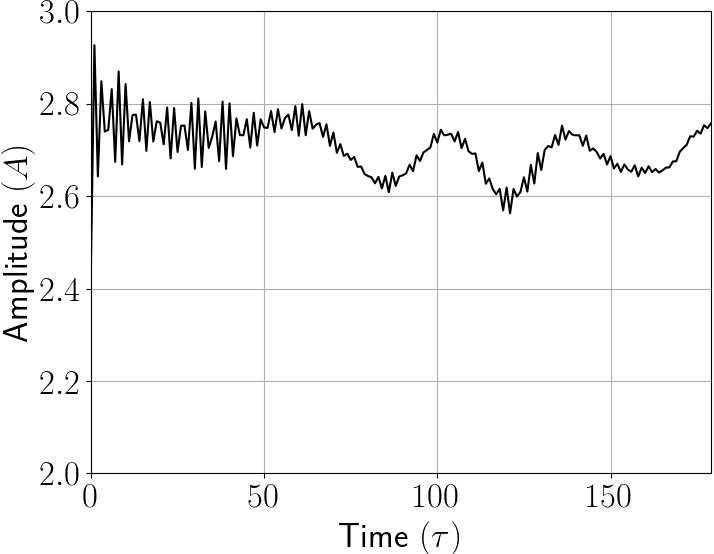}} \hspace{0.2cm}
  \subfloat{\includegraphics[width=0.3\textwidth]{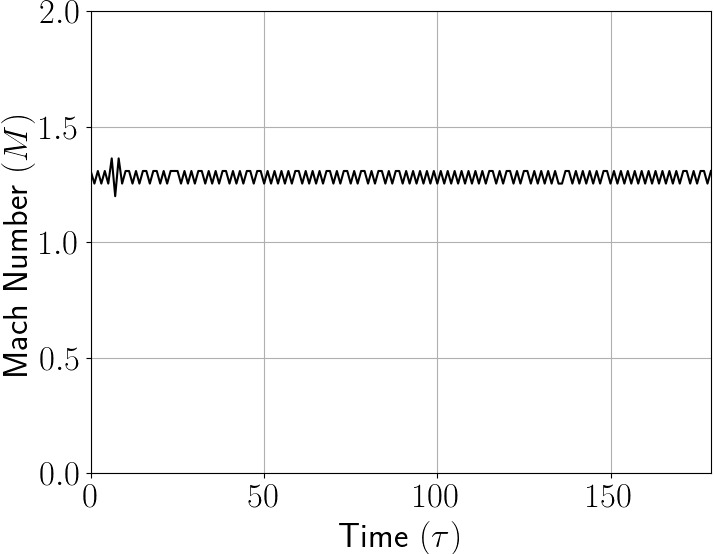}} \hspace{0.2cm}
  \subfloat{\includegraphics[width=0.3\textwidth]{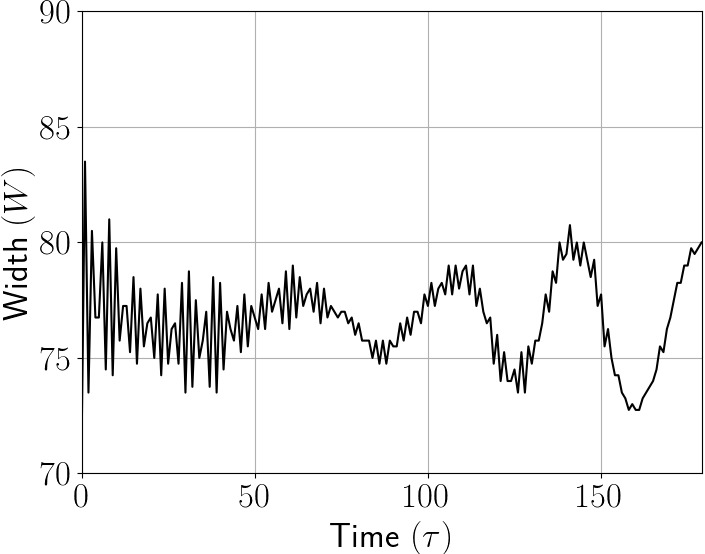}}
  \caption{Temporal evolution of three major features,
  e.g. amplitude, width and Mach number of 
  ion-acoustic solitary waves are presented
  for the case I ($M = 1.28, \beta = 0$). 
  The propagation of the ion-acoustic solitary waves appears as
  slow oscillation in the profile of amplitude and width.
  }
  \label{Fig_M1o28_B0_Max_e_Max_i_soliton_analysis_long}
\end{figure*}

\begin{figure*}
  \subfloat{\includegraphics[width=0.4\textwidth]{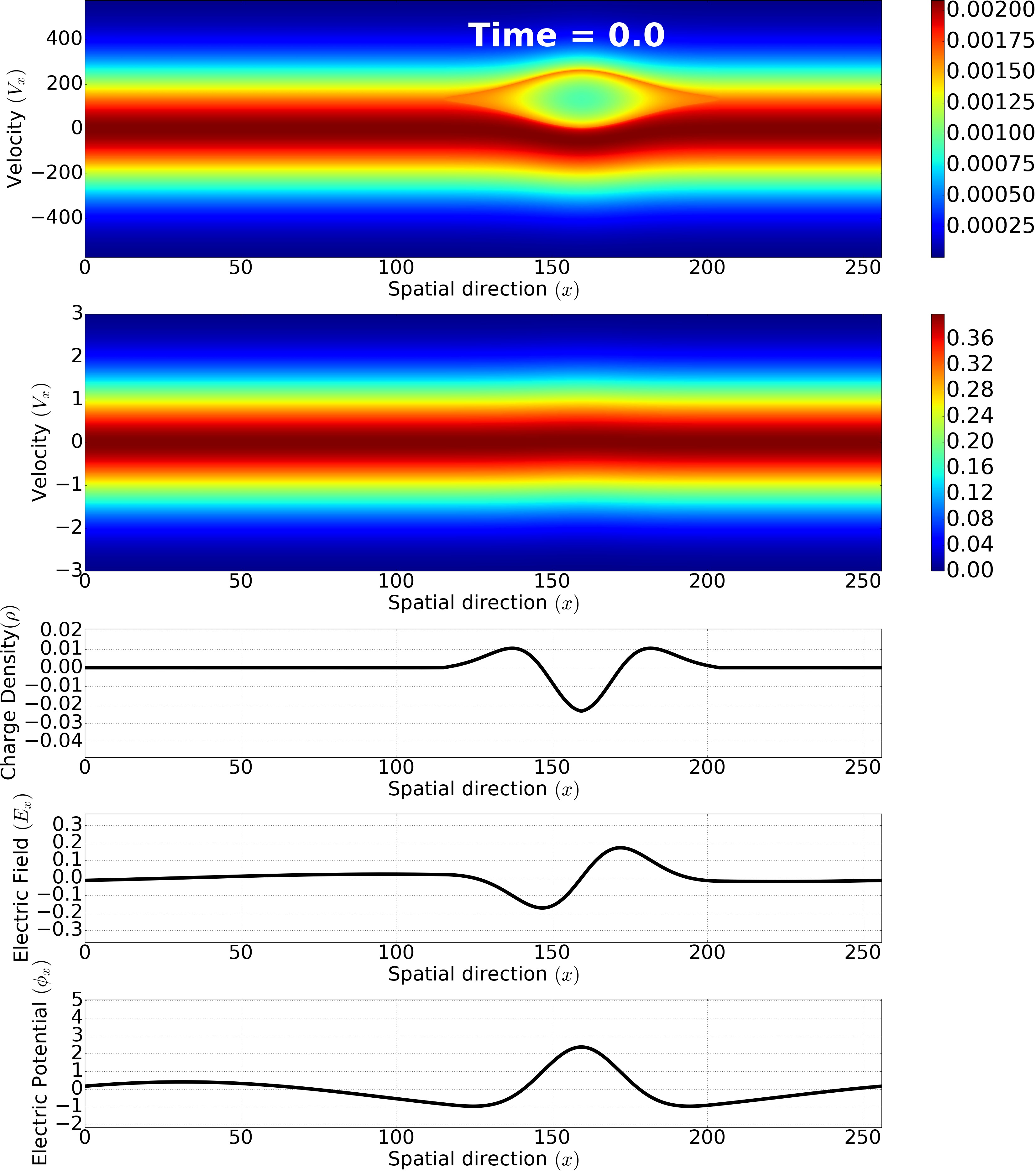}} \hspace{1.0cm}
  \subfloat{\includegraphics[width=0.4\textwidth]{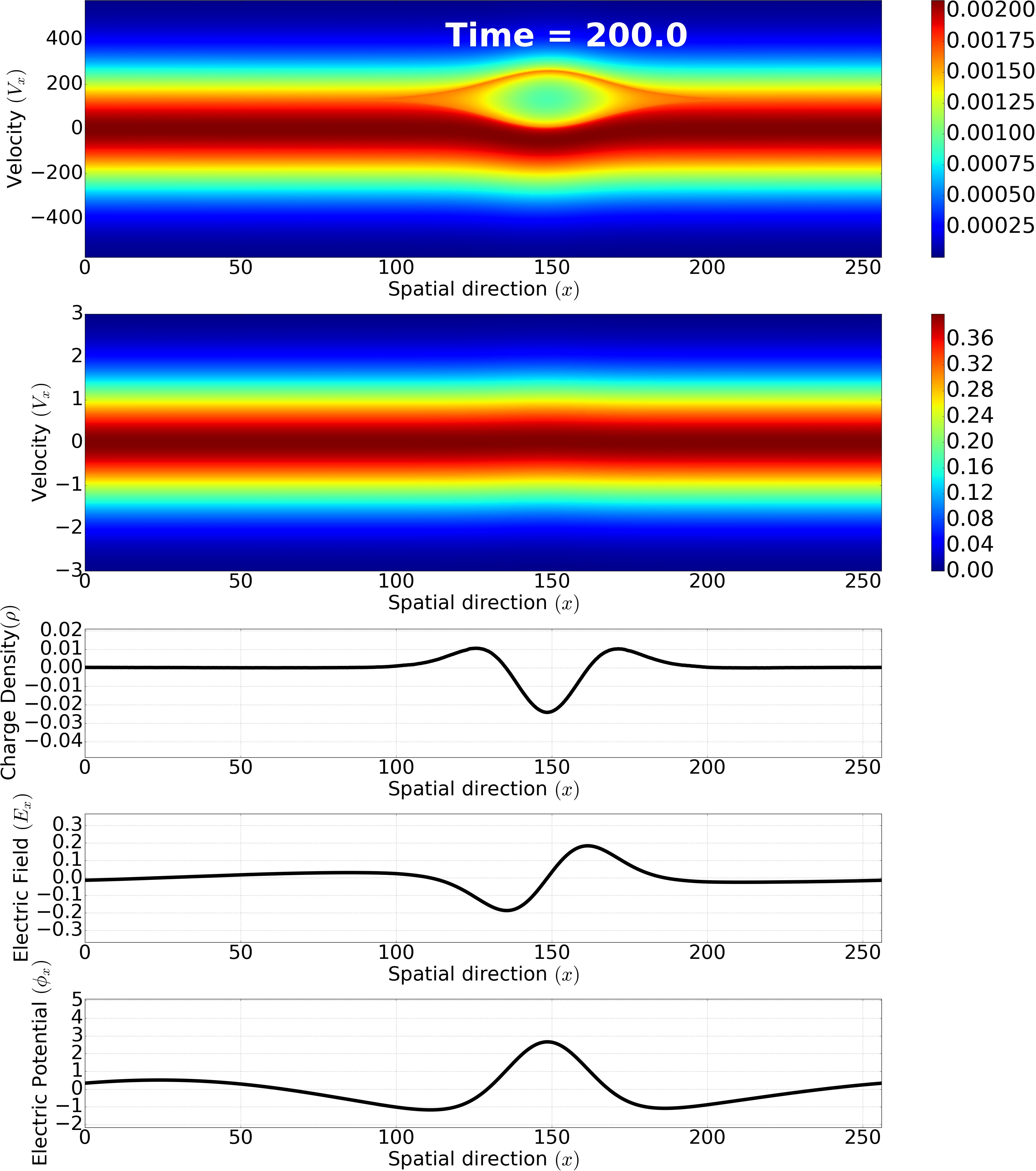}} \\
  \subfloat{\includegraphics[width=0.9\textwidth]{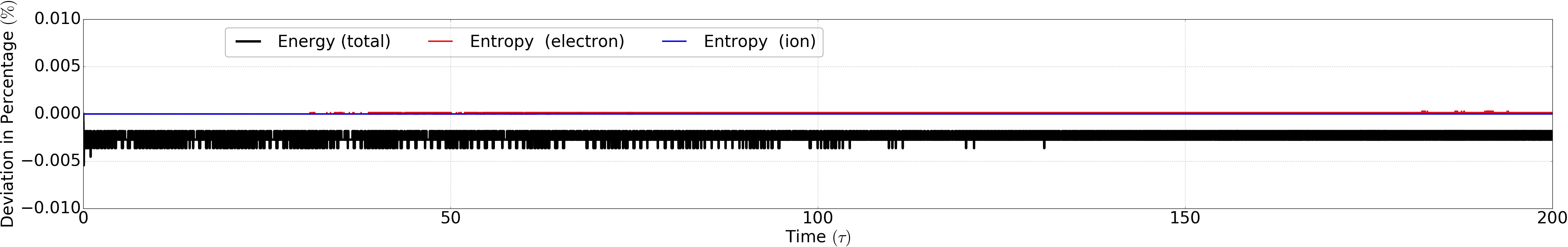}}
  \caption{Simulation results for $M = 29.27$ and $\beta = -2.5$ when distribution functions
	   of both species are Maxwellian in the frame $x-v_s \tau$. 
	   Temporal evolution of electron distribution function (first row),
	   ion distribution function (second row),
	   charge density (third row),  electric field (fourth row) 
	   and electric potential (fifth row) is shown for first and 
	   last time step, e.g. $\tau = 0, 200$. 
	   Conservation of energy and entropy is presented for the simulation which
	   shows deviation below $\%0.1$. 
	   Stability of the Sagdeev solution for a long-time propagation can be observed. }
  \label{Fig_M30_B-2o5_Max_e_Max_i}
\end{figure*}

Furthermore, 
we have presented the temporal evolution of 
the three major characteristics of the solitary waves, 
e.g. amplitude, width and velocity. 
Their stability during propagation reflects 
the stability of solitary waves in the simulation. 
There exist fairly small fluctuations on the average value of 
width and amplitude originating
from the Langmuir and ion acoustic waves,
which are excited due to 
the periodic boundary conditions of the simulation box. 
However, these fluctuations do not have any effect 
on the stability of the solitary waves and 
its Mach number, i.e. velocity, as its main feature. 
The effect of Langmuir wave propagation 
can be observed in the zoom-in set of 
figures dedicated to the early stage 
of the temporal evolution ($0 <\tau< 5$)
in Fig.~\ref{Fig_M1o28_B0_Max_e_Max_i_soliton_analysis_short}.
Comparing this results with the results from study of nonlinear Landau damping of Langmuir mode\cite{abbasi2007vlasov}
shows the similarity between the figures and confirm our interpretation of these oscillations.

The oscillation originating from the propagation of ion-acoustic waves
can be witnessed on the larger time scale and later in the simulation. 
Fig.~\ref{Fig_M1o28_B0_Max_e_Max_i_soliton_analysis_long} shows
these oscillations happening clearly for $\tau>100$ in the two features 
of solitary waves, i.e. width and amplitude. 
However, the effect of these oscillations stays around $5\%$ of the 
average amplitude and width of solitary waves.
Moreover, the Mach number of solitary waves (its velocity) stands 
unaffected by these oscillation,
which shows the survival of the 
solitary waves. 
Note that any change in the overall shape of solitary waves will affect 
its velocity due to the sensitive relation between shape (height and width)
and velocity.

\begin{figure}
  \subfloat{\includegraphics[width=0.5\textwidth]{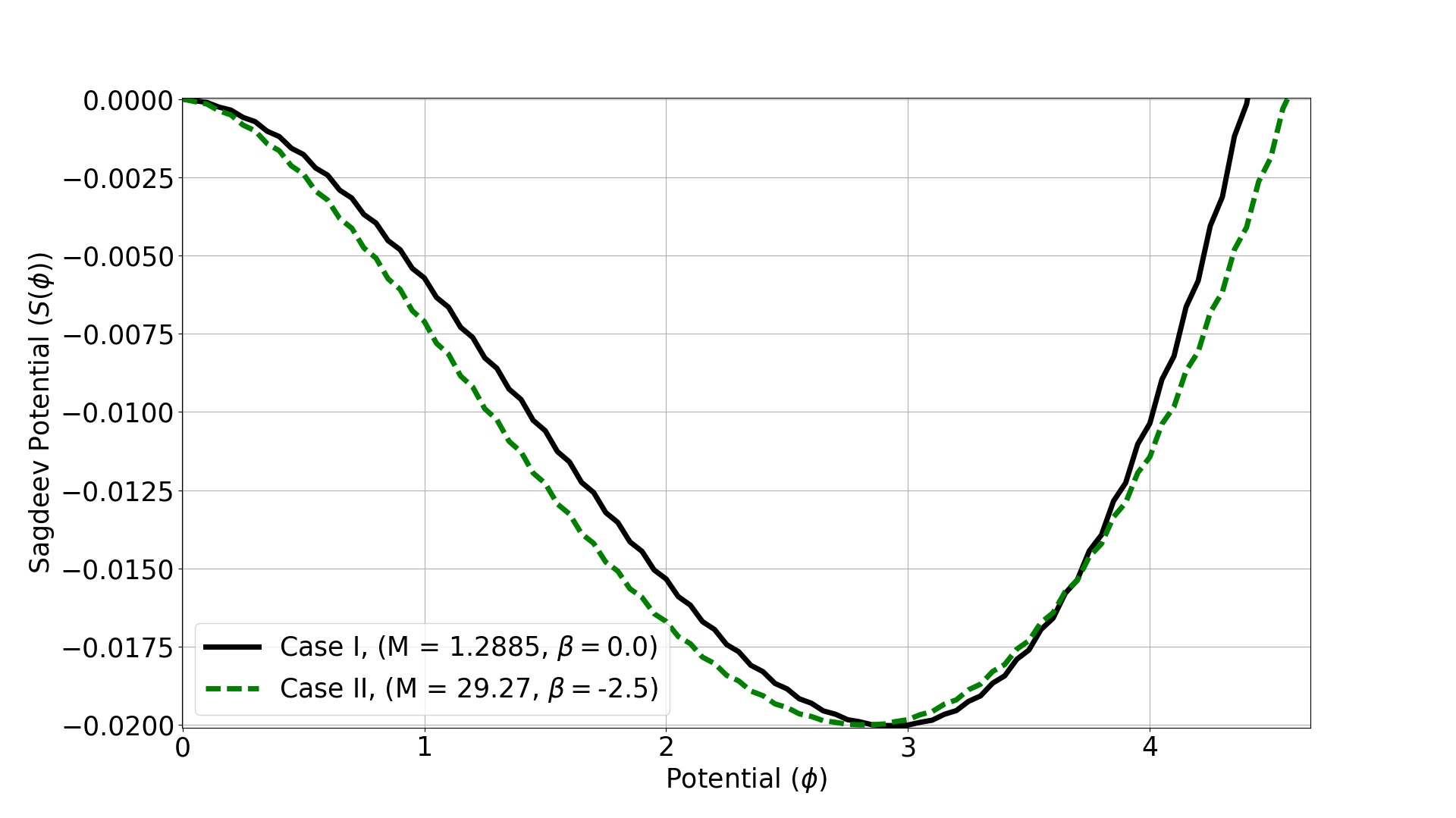}} 
  \caption{The Sagdeev pseudo-potential ($S(\phi)$) for the two cases of
  the study presented when the both species are modeled by the Maxwellian distribution function.}
  \label{Fig_comparison_Sagdeev}
\end{figure}

\begin{figure*}
  \subfloat{\includegraphics[width=0.4\textwidth]{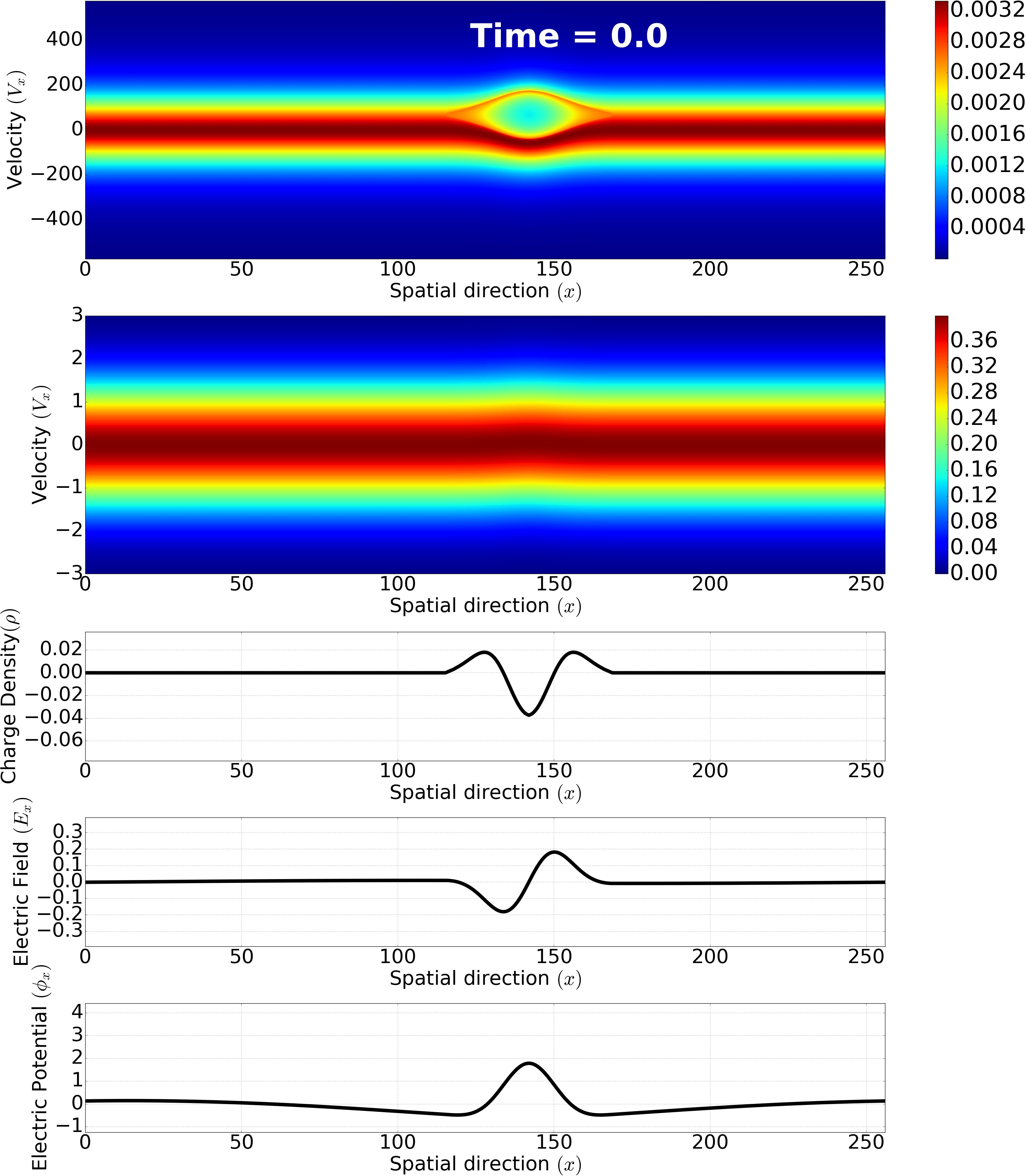}} \hspace{1.0cm}
  \subfloat{\includegraphics[width=0.4\textwidth]{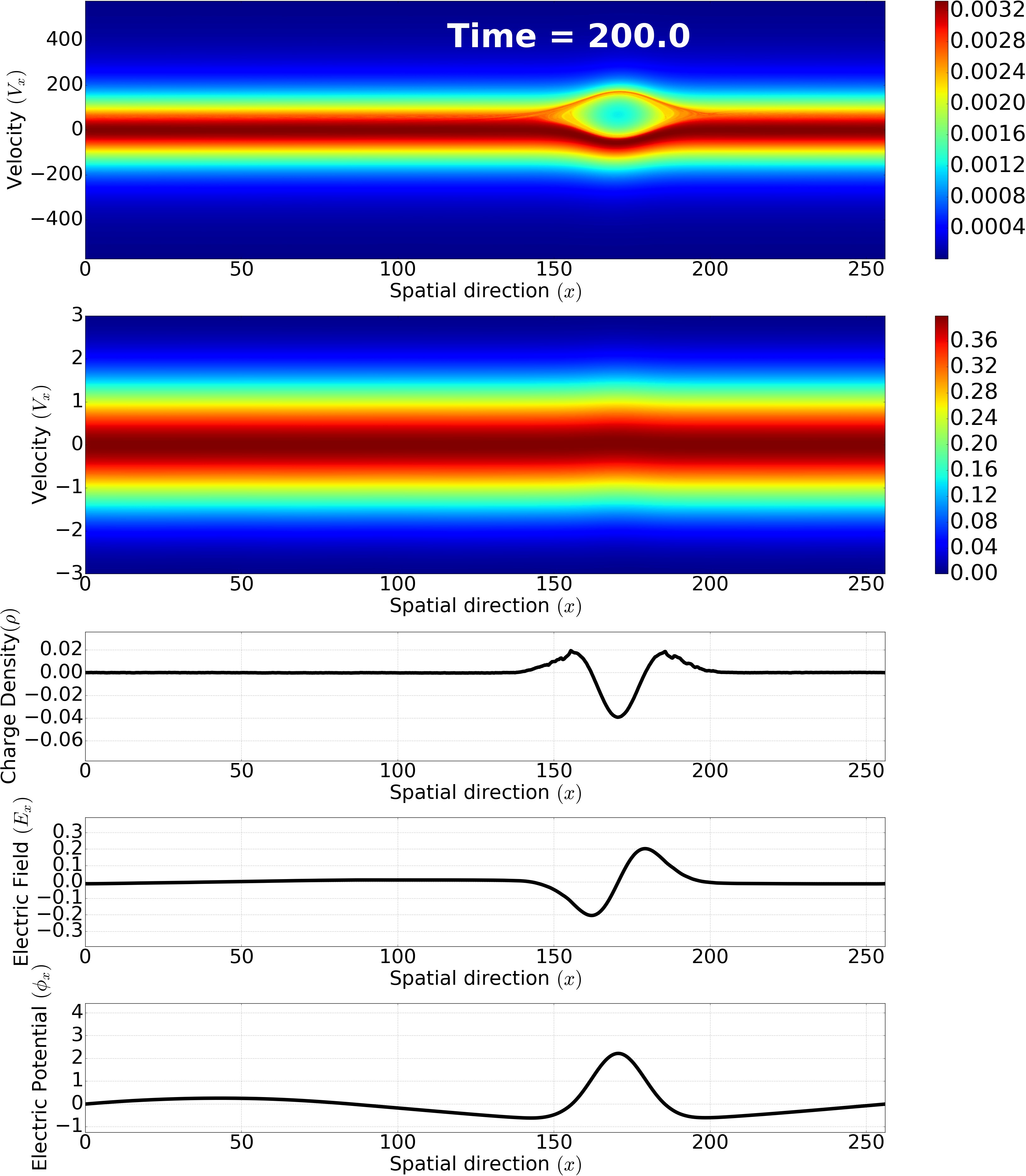}} \\
  \subfloat{\includegraphics[width=0.9\textwidth]{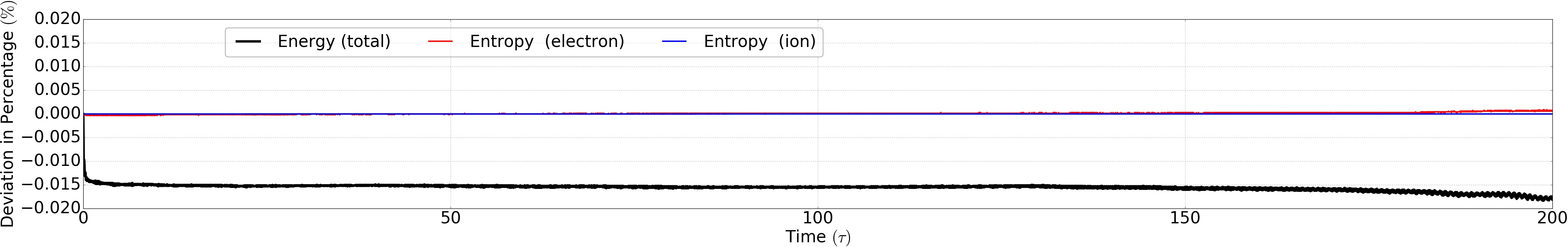}}  
  \caption{Temporal evolution of solitary waves with $M = 15.0, \beta = -5.0$, Maxwellian ions and 
  Kappa distribution function for electron with $\kappa = 2.0$ in presented in the lab frame. Two time steps are shown, e.g. 
  $\tau = 0, 200$. From the top to bottom rows represent, electron phase space,
  phase space of ions,
  charge density, electric field ($E$), electric potential ($\phi$) and 
  conservation of energy and entropy.}
  \label{Fig_M15_B-5_kappa2_e_Max_i}
\end{figure*}

Changing the value of $\beta$ from zero results in the modification of 
the amplitude, width and shape of the Sagdeev solutions. 
We have carried out simulations with different values 
to examine the stability in the parameter space of ($\beta, v_p$).
The simulation results are presented in 
Fig ~\ref{Fig_M30_B-2o5_Max_e_Max_i} for case II ($M = 29.27, \beta = -2.5$).
The parameters are chosen such that ion-acoustic solitary waves 
possess the same amplitude for the electric field as
in Fig.~\ref{Fig_M1o28_B0_Max_e_Max_i},
in order to have the figures comparable. 
The Sagdeev pseudo-potential of the both cases 
are shown in Fig.~\ref{Fig_comparison_Sagdeev}. 
We have also produced a movie for $M = 3.0$ and $\beta = -1$
as well which can be found as multimedia alongside this paper.

Case II presents a much faster solitary wave than case~I with a deeper 
hole in the electron phase space. 
The simulation results prove the stability 
of this fast moving solitary waves
during the long time propagation. 
Based on the analytical fluid approach to the Sagdeev solution (in which the effect
of trapped and reflected particles are ignored) there is a  maximum limit for the 
velocity of ion-acoustic solitary waves ($v_p < 1.6 v_c$ or $M < 1.6$) \cite{Sagdeev, shukla1983solitons}.
In this analytical model, electrons are considered as Boltzmann fluid and ions are considered cold ($T_i \ll T_e$)
and the Sagdeev pseudo-potential reads as follow\cite{shukla1983solitons}:
\begin{equation*}
    S(\phi) = 1 - \exp (\phi) + M^2 \Big[ 1 - \big(1-\frac{2 \phi}{M^2}\big)^{1/2}  \Big].
\end{equation*}
Our simulations reveals that by removing these constrains and considering the kinetic effects, 
solitary waves can move much faster, at least as high as $M = 30$ based on case II of our simulations.

Next, we have employed our fully kinetic simulation approach
for the Kappa distribution function with $\kappa = 2, M = 15, \beta = -5$.
Fig.~\ref{Fig_M15_B-5_kappa2_e_Max_i} shows the stability of the ion-acoustic solitary
waves in this regime. 

  \subsection{Instability of the Sagdeev solutions} \label{SubSec_Results_Instability}  

\begin{figure*}
  \subfloat{\includegraphics[width=0.4\textwidth]{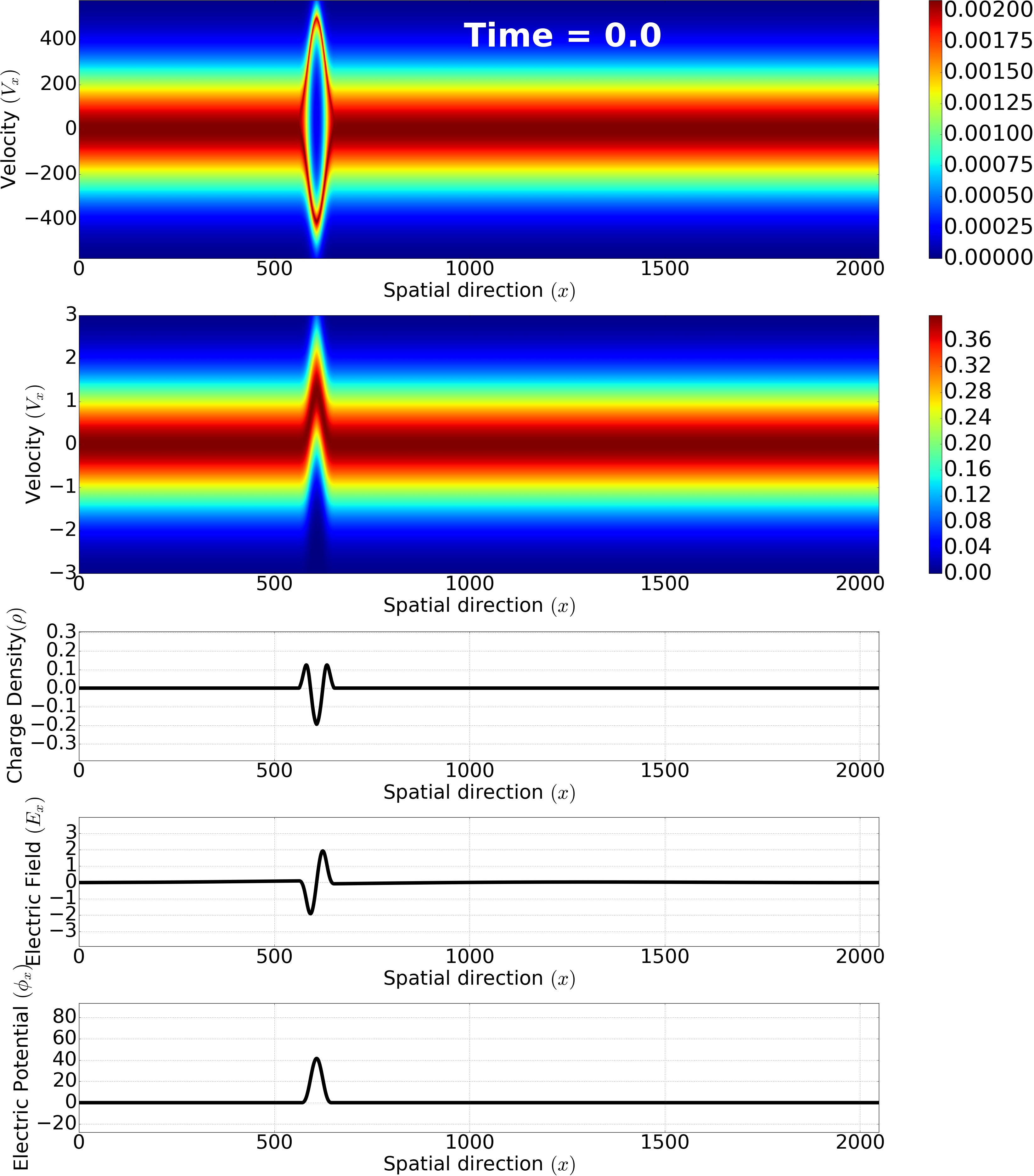}} \hspace{1.0cm}
  \subfloat{\includegraphics[width=0.4\textwidth]{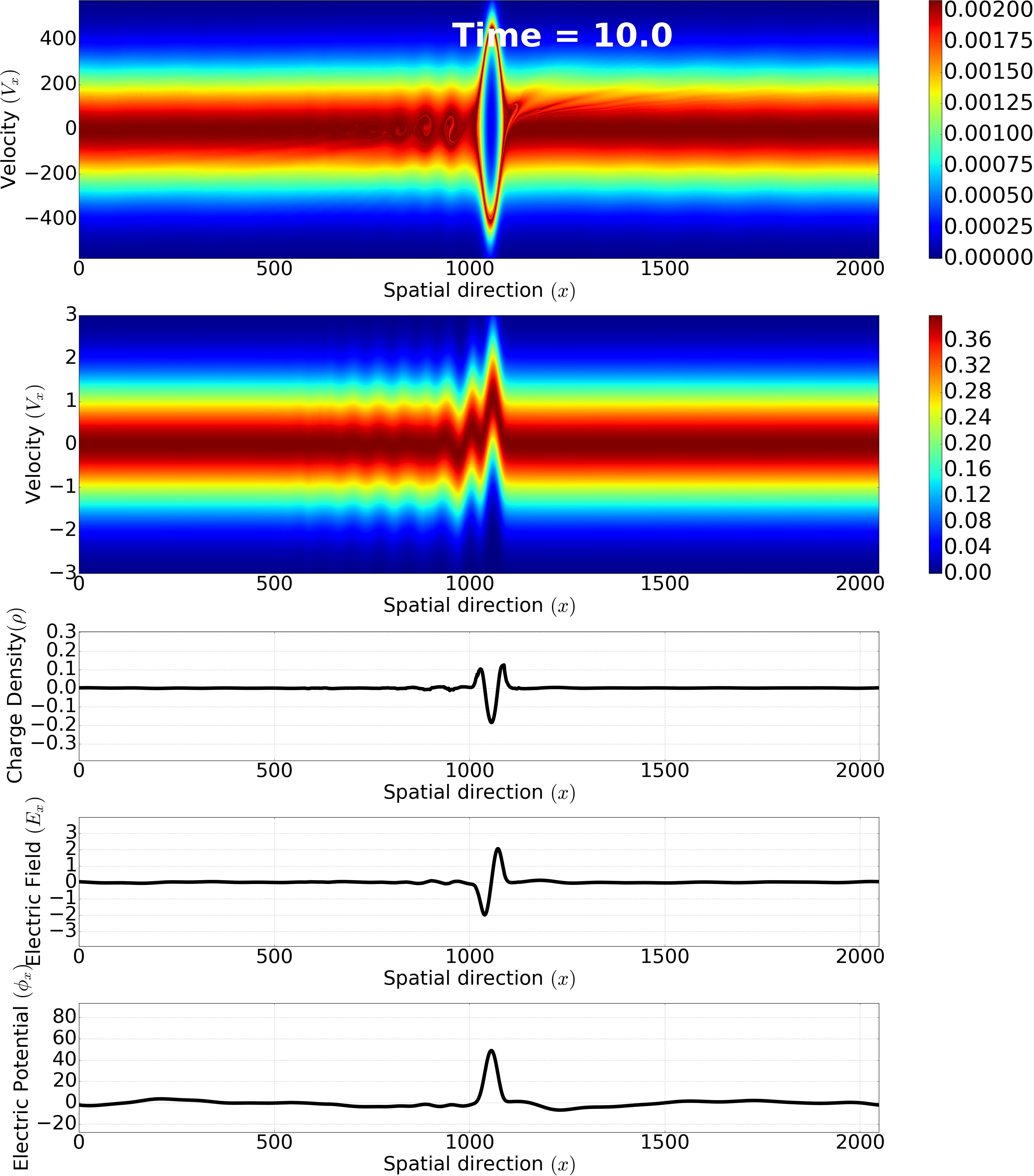}} \\
  \subfloat{\includegraphics[width=0.9\textwidth]{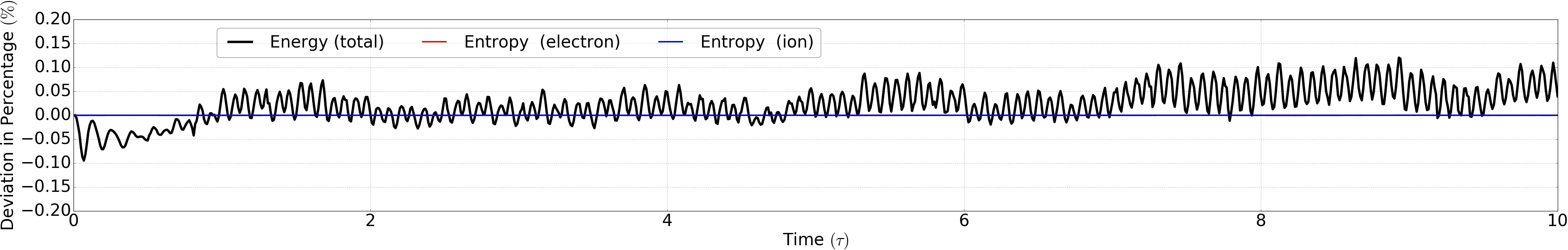}}  
  \caption{Temporal evolution of a Sagdeev solution with $M = 10.0, \beta = -0.8$ 
  with Maxwellian distribution function for both ions and electrons is presented in the lab frame
  for two time steps $\tau = 0, 10$.
  This case presents a fast and unstable Sagdeev solution. 
  From top to bottom, phase space of electrons and ions, charge density, electric field and potential and deviation 
  of energy and entropy are shown.
  }
  \label{Fig_SagdeevDecay_M10B-0o8L2048}
\end{figure*}

\begin{figure*}
  \subfloat{\includegraphics[width=0.4\textwidth]{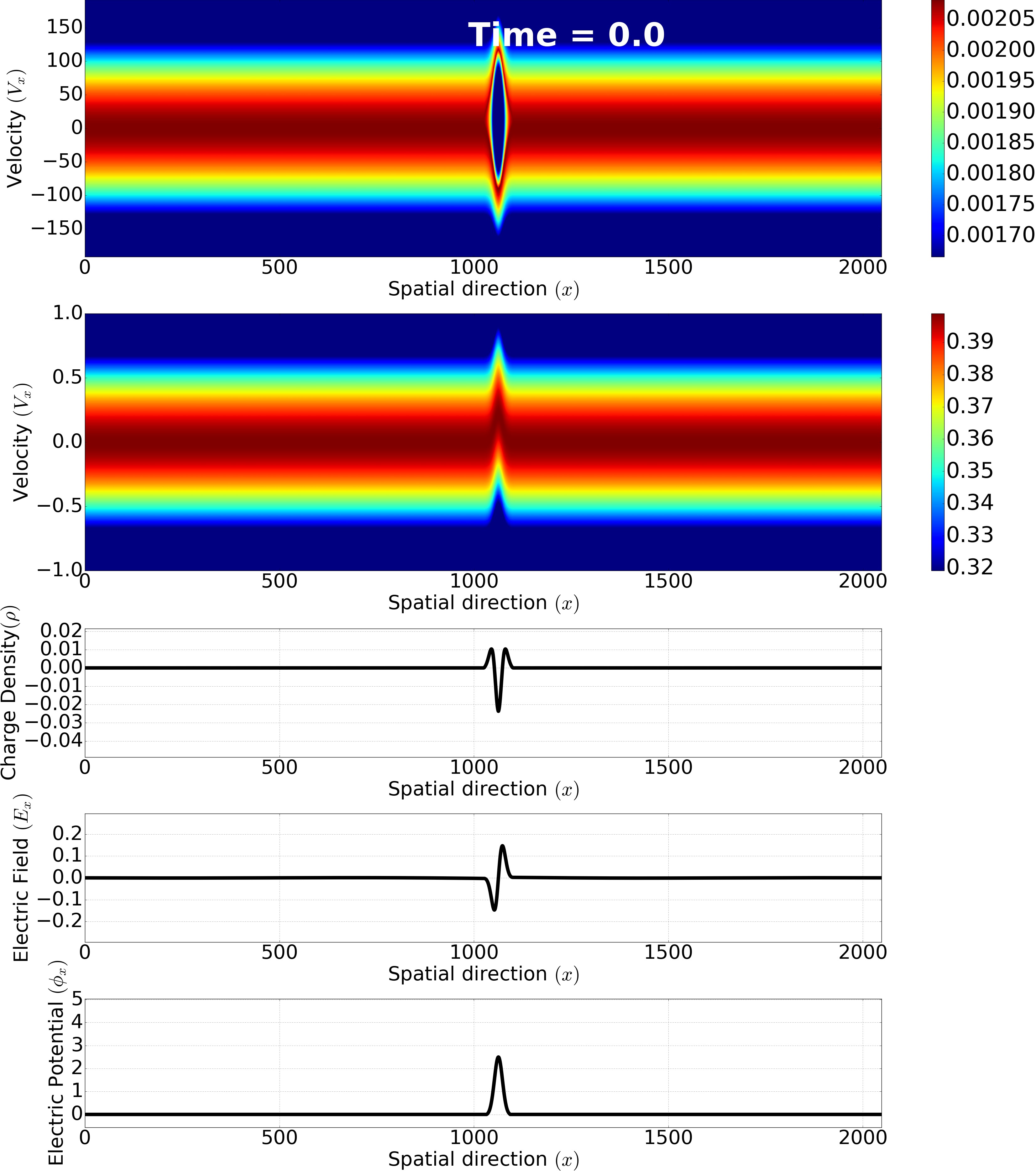}} \hspace{1.0cm}
  \subfloat{\includegraphics[width=0.4\textwidth]{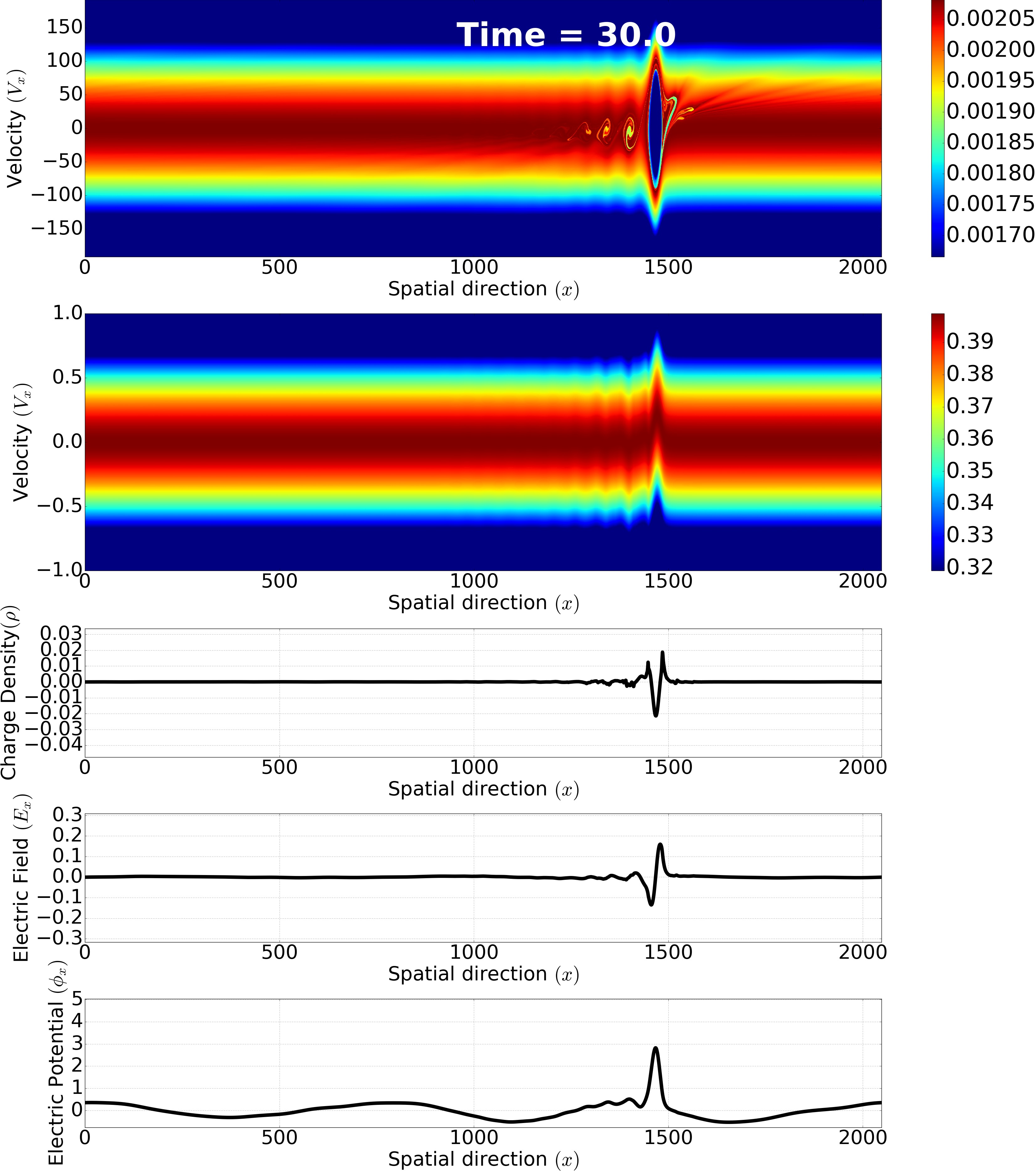}} \\
  \subfloat{\includegraphics[width=0.9\textwidth]{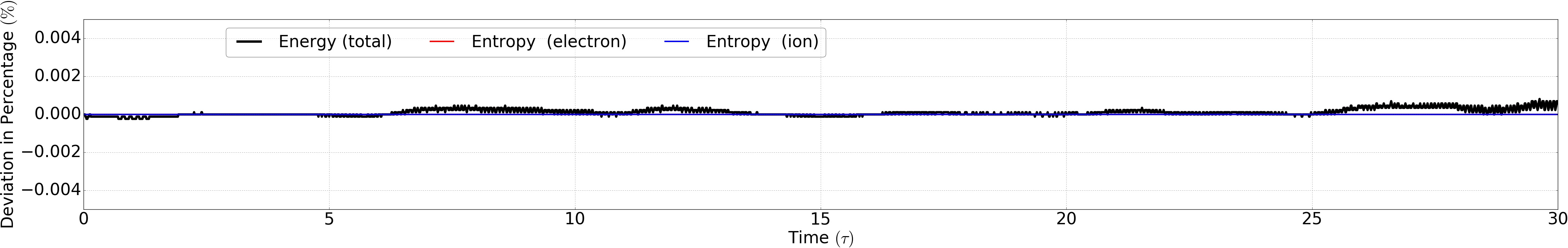}}  
  \caption{Temporal evolution of a Sagdeev solution with $M = 3.0, \beta = -3.5$ 
  with Maxwellian distribution function for both ions and electrons is presented in the lab frame.
  This case presents a slow and unstable Sagdeev solution. 
  Five quantities are shown in two time steps ($\tau = 0, 30$) starting from the top phase space of electrons
  and ions, charge density, electric field and potential. 
  The conservation of energy and entropy (for both ions and electrons) stays below $0.002\%$ shown in the last row.
  }
  \label{Fig_SagdeevDecay_M3B-3o5_L2048}
\end{figure*}

  Instability of Sagdeev solutions has been reported just once (to the best of our knowledge),
  recently in a PIC simulation study by Zhou and Hutchinson \cite{zhou2017plasma}. 
  They have employed a transient approach to seed the electron hole and when it reached 
  its steady-state form, it has been pushed down to examine the effect of velocity on its stability.
  It is concluded from their simulations and the theory of ``hole kinematics''\cite{hutchinson2016plasma} that slow electron holes
  decay into ion-acoustic waves due to an oscillatory velocity instability.
  However, we have observed in our simulations an instability in Sagdeev solutions for even high velocity solutions.
  Fig.\ref{Fig_SagdeevDecay_M10B-0o8L2048}  presents the results for a case of simulation with $M=10, \beta=-0.8$.
  Sagdeev solutions with $M=10, \beta = -2, -3, -5, -7$ are tested and they appear to be strongly stable. 
  The same process of instability has been observed for slower Sagdeev solutions in our simulations.
  Fig.\ref{Fig_SagdeevDecay_M3B-3o5_L2048} shows an example of such case with $M=3, \beta=-3.5$.
  Sagdeev solutions of $M=3.0$ with $\beta = -4, -5$ stay stable while $\beta = -3, -2, -1$ shows the same type of instability
  as $\beta = -3.5$ with minor changes in time scales.
  
\begin{figure*}
  \subfloat{\includegraphics[width=0.9\textwidth]{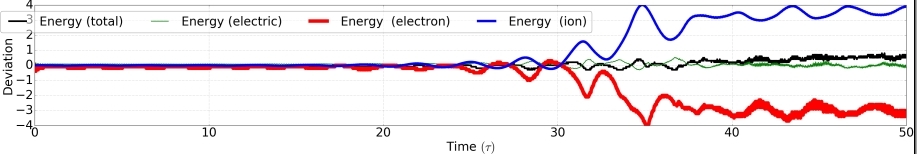}} \hspace{1.0cm}
  \caption{Temporal evolution of energy of each species alongside the electric and total energy for the case $M = 3.0, \beta = -3.5$ 
  with Maxwellian distribution function for both ions and electrons is presented. 
  The deviation is calculated by $X(\tau) - X(0)$.
  }
  \label{Fig_SagdeevDecay_M3B-3o5_L512_Energy_Channel}
\end{figure*}
\begin{figure*}
  \subfloat{\includegraphics[width=0.9\textwidth]{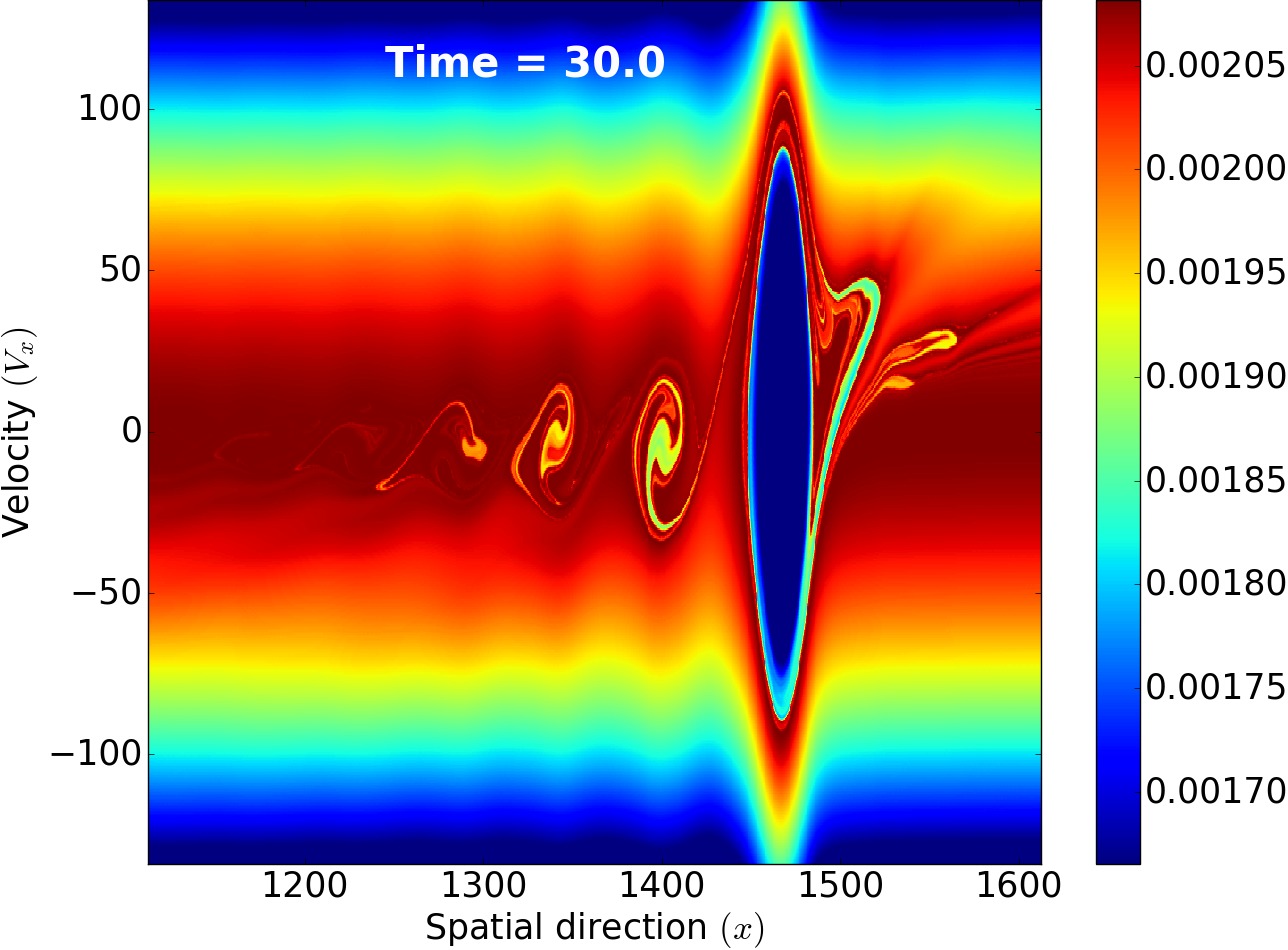}} \hspace{1.0cm}
  \caption{Process of detrapping is shown in details for $M=3.0, \beta = -3.5$ at time step $\tau = 30$ 
  (see Fig.\ref{Fig_SagdeevDecay_M3B-3o5_L2048}).
  The instability causes the electron hole to lose some parts of its trapped population which appear as
  gradually decreasing holes in the wake of its propagation.}
  \label{Fig_SagdeevDecay_M3B-3o5_L2048_T30_Details_Detrapping}
\end{figure*}

  The process of instability observed in our simulation appears 
  as an energy exchange between electrons and ions.
  Fig.\ref{Fig_SagdeevDecay_M3B-3o5_L512_Energy_Channel} demonstrates 
  the growth of kinetic energy of ions in the expense of 
  electrons' kinetic energy, while the change in electric energy 
  is negligible. 
  This process has been predicted to exist by the theory of ``hole kinematics''\cite{zhou2017plasma},
  however this is the first time to report it. 
  Furthermore, the instability causes the Sagdeev solution to emit a trail of ion-acoustic wave packet in 
  the wake of its propagation. 
  Since Sagdeev solutions are faster than the ion-acoustic speed,
  hence the Sagdeev solution stays a head of the ion-acoustic wave-packet which appear
  as gradually decreasing holes behind it (see Fig.\ref{Fig_SagdeevDecay_M3B-3o5_L2048_T30_Details_Detrapping}).
  In other words, some parts of trapped population in the electron holes become detrapped due to the loss of 
  amplitude of the potential well.

  Our simulations clearly show that the trapping parameter play a crucial role in the 
  stability of the Sagdeev solution. 
  For any specific velocity (Mach number), there exists a threshold in the 
  trapping parameter ($\beta_c$) which separates the stable solutions from 
  unstable ones. 
  In cases reported here, it is $\beta_c = -1, -3.5$ for $M = 10, 3$ respectively.

A full study of this phenomena 
will be communicated in the future reports. 
However, we are able to claim
that the fully kinetic Sagdeev solutions 
in the ion-acoustic regime are not guaranteed to 
be stable.
This can affect the implication of the Sagdeev method 
in more complex scenarios such as multi-species plasmas. 
What our simulation reveals is that the Sagdeev solutions should not 
be automatically considered as solitary waves. 
Their stability should be established either by studying their behavior
in long-term simulations or through theoretical approach such as
``hole kinematics''\cite{hutchinson2016plasma}.


\section{Conclusions} \label{Sec_Conclusions}
The nonlinear solutions of the Sagdeev pseudo-potential approach 
to the ion-electron plasmas are studied in the fully-kinetic ion-acoustic regime.
This research stands as the first attempt to verify 
Sagdeev solutions as solitary waves in the kinetic regime 
(to the best of our knowledge).
Our simulation results show both stability and instability of Sagdeev solutions 
depending on the combination of parameters, i.e. velocity $M$ and trapped parameter $\beta$.
Our results demonstrate that in the presence of numerical noise 
in a long-time simulation, some of Sagdeev solutions stay stable.
This at least can prove that
they are locally stable nonlinear modes of the system, hence are solitary waves.

Moreover, our simulations show the instability of Sagdeev solutions for both slow and fast 
moving ones. 
Instability grows from small numerical noises and causes 
the nonlinear pulse to lose its trapped population by
emitting ion-acoustic waves packets on its propagation trail. 
We have found that there exists a clear threshold in trapping parameter ($\beta_c$)
for each velocity ($M$) which separates the stable solutions from unstable ones. 

In case of stable solutions,
verification has been presented in the three types of plots 
reflecting different aspects of the dynamics 
in both phase space and real space. 
In the real space, the temporal evolution of three quantities, 
namely charge density, electric field and potential are analyzed
for any instability. 
Furthermore, the features of solitary waves such as 
velocity, width and amplitude are closely monitored. 
Although width and amplitude show oscillations due to
propagation of low (Langmuir) and high (ion-acoustic) modes 
(related to the use of periodic boundary conditions), 
velocity stands unchanged which reflects the stability of Sagdeev solutions.
In the phase space, 
the evolution of both distribution function and 
energy trajectories are presented and examined
for any deviation from initial shape of the solitary waves.

Furthermore, high-speed ion-acoustic solitary waves ($M=30$) are presented
which demonstrates the existence of solitary waves out of 
existence regime proposed by fluid model.
This shows the great impact of the kinetic effects on the existence regime, 
ignored in the fluid model.
Kinetic effects such as trapping and reflection modifies the plasma such that
large Mach number Sagdeev solutions (solitary waves) are possible.

In order to validate our simulation code, 
we have chosen the nonlinear problem of the effect of ion dynamics
on standing electron holes as a benchmarking test. 
The results confirms the theoretical predictions and 
shows the robustness of the VHS (Vlasov-Hybrid Simulation) method
in following the temporal evolution of the phase space in nonlinear stage.

Furthermore, we have presented the conservation of both energy and 
entropy for all the simulations discussed in this study. 
Deviation from the initial value for all the conserved quantities
reported here, stays below $0.1\%$

\section{Supplementary Material} \label{Sec_Supp_Material}
Temporal evolution of ion acoustic solitary waves for a movie for $M = 3.0$  and $\beta=−1$
is presented for different variables including:
electron distribution function (first row), 
electron number density(second row),
ion distribution function (third row), 
ion number density (fourth row), 
charge density (fifth row),
electric field (sixth row) 
and electric potential (seventh row) is shown.
Conservation of energy (eighth row) and entropy (ninth row)
is presented for the simulation which shows deviation below $\%0.1$.
The stability of the Sagdeev solution for a long-time propagation can be observed.

\section{Outlook} \label{Sec_outlook}
This self-consistent kinetic approach provides an 
accurate and strong method to study the solitary waves 
in different environments such as multi-species plasmas with 
different types of distribution functions. 
It is being developed and adopted to study structures 
such as super-solitons\cite{Verheest2013} and 
negative solitons in electronegative plasmas 
(consist of positive and negative ions and electrons)
\cite{sheridan1999excitation} and 
other types of multi-species plasmas
and has revealed some interesting phenomenon, which will be
discussed in forthcoming publications.

The work on examining the relationship between stability of Sagdeev solutions and 
other parameters such as mass and temperature ratio is underway.
We are also examining the dependency of 
the critical trapping parameter ($\beta_c$) on the velocity ($M$).

\acknowledgments
G. Brodin and S. M. Hosseini Jenab would like to acknowledgment financial support 
by the Swedish Research Council, grant number 2016-03806.
This work is based upon research supported by the National Research Foundation (NRF)
and Department of Science and Technology (DST) from Republic of South Africa.
Any opinion, findings and conclusions or recommendations expressed in this 
material are those of the authors and therefore the NRF and DST do not accept 
any liability in regard thereto.


%

\end{document}